\documentclass[journal,twocolumn]{IEEEtran}  

\usepackage{amsmath}
\usepackage{amssymb}
\usepackage{graphicx}
\usepackage[usenames, dvipsnames]{xcolor}
\usepackage{tikz}
\usetikzlibrary{spy,arrows}
\usepackage{navigator}

\usepackage{url}
\usepackage{array}
\usepackage{placeins}
\usepackage{txfonts}

\usepackage{afterpage}

\graphicspath{{pix/}}

\usepackage{natbib}

\usepackage{hyperref}

\begin{document} 

\bstctlcite{IEEEexample:BSTcontrol}


    \title{Simulated {JWST} datasets for multispectral and hyperspectral image fusion}

%
%

  \author{Claire Guilloteau$^{1,2}$, Thomas Oberlin$^{3}$, Olivier Bern\'{e}$^{2}$ and Nicolas Dobigeon$^{1,4}$\\[0.5cm]\small
$^{1}$ University of Toulouse, IRIT/INP-ENSEEIHT, Toulouse, France \\
(e-mail: \{Claire.Guilloteau, Nicolas.Dobigeon\}@enseeiht.fr).\\
$^{2}$ University of Toulouse, Institut de Recherche en Astrophysique et Plan\'etologie (IRAP), Toulouse, France \\(e-mail:
    \{Claire.Guilloteau,Olivier.Berne\}@irap.omp.eu)\\
$^{3}$  University of Toulouse, ISAE-SUPAERO, Toulouse, France \\(e-mail: Thomas.Oberlin@isae-supaero.fr)\\
$^{4}$ Institut Universitaire de France (IUF)}

   \maketitle
 
  \begin{abstract}
   The James Webb Space Telescope (JWST) will provide multispectral and  hyperspectral infrared images of a large number of astrophysical scenes. Multispectral images will have the highest angular resolution, while hyperspectral images (e.g., with integral field unit spectrometers) will provide the best spectral resolution.
   This paper aims at  providing a comprehensive framework to generate an astrophysical scene and to simulate realistic hyperspectral and multispectral data acquired by two JWST instruments, namely NIRCam Imager and NIRSpec IFU. We want to show that this simulation framework can be resorted to  assess the benefits of fusing these images to recover an image of high spatial and spectral resolutions.
  To do so, we create a synthetic scene associated with a canonical infrared source, the Orion Bar. 
  This scene combines pre-existing modelled spectra provided by the JWST Early Release Science Program 1288 and real high resolution spatial maps from the Hubble space and ALMA telescopes. We develop forward models including corresponding noises for the two JWST instruments based on their technical designs and physical features. JWST observations are then simulated by applying the forward models to the aforementioned synthetic scene. We test a dedicated fusion algorithm we developed on these simulated observations.
   We show the fusion process reconstructs the high spatio-spectral resolution scene with a good accuracy on most areas, and we identify some limitations of the method to be tackled in future works.
   The synthetic scene and observations presented in the paper are made publicly available and can be used for instance to evaluate instrument models (aboard the JWST or on the ground), pipelines, or more sophisticated  algorithms dedicated to JWST data analysis. Besides, fusion methods such as the one presented in this paper are shown to be promising tools to fully exploit the unprecedented capabilities of the JWST.
  \end{abstract}


%

\section{\label{intro}Introduction}

The James Webb Space Telescope (JWST) is an international collaboration space observatory involving NASA, European Space Agency (ESA) and Canadian Space Agency (CSA) and is planed to be launched in 2021 \citep{Gardner2006}.
The four embedded instruments, the Near InfraRed Camera (NIRCam, \citealt{Rieke2005}), the Near InfraRed Specrtograph (NIRSpec, \citealt{Bagnasco2007}), the Near InfraRed Imager and Slitless Spectrograph (NIRISS, \citealt{Doyon2012}) and the Mid InfraRed Instrument (MIRI, \citealt{Rieke2015}), will cover the infrared wavelength range between $0.6$ to $28$ microns with an unprecedented sensitivity. 
The JWST will enable research on every epoch of the history of the Universe, from the end of the dark ages 
to recent galaxy evolution, star and planet formation. The scientific focuses of the JWST range from first light and reionization to planetary systems and the origins of life, through galaxies and protoplanetary systems. 
The JWST mission  will observe with imagers and spectrographs. The imagers of NIRCam and MIRI will provide multispectral images (with low spectral resolution) on wide fields of view (with high spatial resolution) while the spectrometers Integral Field Units (IFU) of NIRSpec and MIRI will provide hyperspectral images (with high spectral resolution) on small fields of view (with low spatial resolution).
The aim of the present study is to assess the possible benefits of combining complementary observations, i.e., multispectral and hyperspectral data, of the same astrophysical scene to reconstruct an image of high spatial \emph{and} spectral resolutions. If successful, such a method
would provide IFU spectroscopy with the spatial resolution of the imagers. For the near infrared range, which is the focus of this paper this corresponds to an improvement by a factor of $\sim 3$ of the angular resolution of NIRSpec IFU cubes, using the NIRCam images. Practically, this implies the possibility to derive integrated maps in spectral features (e.g. H recombination lines, ions, H$_2$) at the resolution of NIRCam and at wavelengths where this latter instrument does not have any filter over the NIRSpec field of view. This may prove useful to derive high angular resolution maps of the local physical conditions which requires the use of a combination of lines.  

In the geoscience and remote sensing literature, the objective described here-above is usually referred to as ``image fusion''. 
State-of-the-art fusion methods are based on an inverse problem formulation, consisting in minimizing a data fidelity term complemented by a regularization term \citep{Wei2015,Simoes2015}. The data fidelity term is derived from a forward model of the observation instruments. The regularization term can be interpreted as a prior information on the fused image. \citet{Simoes2015} proposed a total-variation based prior and an iterative solving while \citet{Wei2015} introduced a fast resolution by defining an explicit solution based on a Sylvester equation, thus substantially decreasing the computational complexity.
Alternatively, \citet{Yokoya2012} proposed a method based on spectral unmixing called coupled non-negative matrix Factorization (CNMF). Elementary spectral signatures, usually referred to as endmembers, and their relative proportions in the image pixels are estimated by an alternating NMF on the hyperspectral and multispectral images related through the observation model.  In the particular context of JWST astronomical imaging, the first challenge of data fusion is due to the very large scale of the fused data, considerably larger than the typical sizes of data encountered in Earth observation. Indeed, a high spatio-spectral fused image in remote sensing is composed of approximately a few ten of thousands pixels and at most a few hundred of spectral points versus a few ten of thousands pixels and a few thousand spectral points for a high spatio-spectral fused image in astronomical imaging. Moreover, another issue in astronomical images fusion is the complexity of observation instruments. Some specificities, such as the spectral variability of point spread functions (PSFs), cannot be neglected because of the large wavelength range of the observed data. Therefore, remote sensing data fusion methods are not appropriate to fuse astronomical observation images. To address these issues, we discuss the relevance of a new fusion method specifically designed to handle JWST measurements.

To assess the relevance of fusing hyperspectral and multispectral data provided by the JWST instruments, a dedicated comprehensive framework is required, in the same spirit as the celebrated protocol proposed by \citet{Wald2005} to evaluate the performance of remote sensing fusion algorithms. This framework mainly relies on a reference image of high spatial and high spectral resolutions and the instrumental responses applied to this image to generate simulated observations. In the context of the JWST, the use of  simulated observations with reference ground truth image is inevitable since, first, real data is not available yet, and second, because only synthetic data allow the algorithm performances to be quantified. Thus this paper aims at deriving an experimental protocol to evaluate the performance of fusion algorithms for JWST measurements. In the current study, the reference image of high spectral and high spatial resolutions, referred to as {\it synthetic scene} hereafter, has been synthetically created to fit the expected physical properties of a photodissociation region (PDR, see definition in Sect.~\ref{pdr}), covering a 31 $\times$ 31 arcsec$^2$ field of view (FOV) between 0.7 and 28.5 $\mu$m. This choice has been driven by our involvement in the JWST Early Release Science (ERS) program ``Radiative Feedback from Massive Stars as Traced by Multiband Imaging and Spectroscopic Mosaics'' lead by \cite{Berne2017} and hereafter referred to as ERS 1288\footnote{\url{www.jwst-ism.org}}, following the ID given by Space Telescope Science Institute (STScI). This choice is also motivated by our past expertise on this type of astrophysical source. However one should keep in mind that the proposed simulation protocol and fusion method may in principle be applied to any kind of dataset, with any type of source. Besides, it is worth noting that the simulation of the hyperspectral and multispectral JWST data associated with this synthetic scene is much more complex than the forward models involved in the Wald's protocol mainly due to the specificities of the instruments mentioned above.

The paper is organized as follows. Sect.~\ref{pdr} describes the specific structure of photodissociation regions. Next, in Sect.~\ref{synthpdr}, we create a synthetic spatio-spectral infrared PDR scene located in the Orion Bar with one spectral dimension (from 0.7 to 28 microns) and two spatial dimensions (each one $\sim$ 30 arcsec wide or high). 
In Sect.~\ref{formod}, we properly define the forward models associated with the NIRCam Imager and NIRSpec IFU instruments. These are mathematical descriptions of the light path through the telescope and the instrument and include specificities such as wavelength-dependant PSFs, correlated noise, spatial sub-sampling, among others. We apply these forward models to the PDR synthetic scene, to produce simulated NIRCam Imager and NIRSpec IFU near-infrared observations (0.7-5 $\mu$m) of the Orion Bar PDR. Finally, in Sect.~\ref{expe} we perform symmetric data fusion between NIRCam Imager short wavelength (SW) channel (0.7-2.35 $\mu$m) and NIRSpec IFU simulated data to qualify the fused high spatio-spectral resolution image, and we evaluate the performance of this fusion scheme.


\section{\label{pdr}Photodissociation regions}


The present paper focuses on a synthetic scene of a PDR. We therefore provide in the following section the general aspects of the concept of a PDR. 

In the interstellar medium, photons from massive stars affect matter, which is found to be either ionized, atomic or molecular, each phase with different temperature and density. Transition regions between molecular clouds and ionized regions (H$_\textsc{II}$) are referred to as PDR \citep{Tielens1985}. This concept of PDR is applicable to many regions in the Universe, such as the surface of Protoplanetary disks \citep{Adams2004, Gorti2008, Champion2017}, as well as planetary nebulae (see e.g. \citealt{Bernard2005, Cox2016}). More broadly, star-forming and planet-forming regions can be studied as PDRs (see for instance \citet{Tielens2005, Goicoechea2016, Joblin2018}), or even starburst galaxies \cite{Fuente2005}. 
Observations of nearby and spatially extended PDRs are essential to characterize, as accurately as possible, their physical and chemical properties, and to benchmark models. This can be done using spatio-spectral maps of PDRs in the main  fine-structure cooling lines of ions and atoms (in particular C$^+$ and O), or molecules such as $\mathrm{H}_2$ \citep{Habart2011,Bron2014}, CO \citep{Joblin2018}, or HCO$^+$ \citep{Goicoechea2016}. From such observations and using PDR models (see a comparision of PDR models by \citet{Roellig2007}), temperature, electronic density and pressure gradient maps with high spatial resolution can be extracted. Observations of rotational and rovibrational lines of H$_2$ can also give clues about H$_2$ formation processes \citep{Bron2014}. Finally, there are numerous studies dedicated to the evolution and photochemistry of Polycyclic Aromatic Hydrocabons (PAHs), a family of large carbonaceous molecules which is ubiquitous in the universe, that are conducted in PDRs, see, e.g., recent examples by \cite{Berne2015,Peeters2017}.

\begin{figure*}[p]
    \centering
    \begin{tikzpicture}
    \draw [fill=red!15] (0,0) rectangle (4,5);
    \shade[left color=red!20, right color=green!15] (4,0) rectangle (4.5,5);
    \draw (4,0) rectangle (4.5,5);
    \draw [fill=green!15](4.5,0) rectangle (7,5);
    \shade[left color=green!15, right color=blue!20] (7,0) rectangle (7.5,5);
    \draw (7,0) rectangle (7.5,5);
    \draw [fill=blue!15](7.5,0) rectangle (10,5);
    
    \draw[<-,violet] (10.3,3) -- (10.8,3);
    \draw[<-,violet] (10.3,1) -- (10.8,1);
    \draw[<-,violet] (10.3,2) -- (10.8,2);
    \draw[<-,violet] (10.3,4) -- (10.8,4);
    \draw[<-,violet] (10.3,0.5) -- (10.8,0.5);
    \draw[<-,violet] (10.3,2.5) -- (10.8,2.5);
    \draw[<-,violet] (10.3,3.5) -- (10.8,3.5);
    \draw[<-,violet] (10.3,1.5) -- (10.8,1.5);
    \draw[<-,violet] (10.3,4.5) -- (10.8,4.5);
    \draw[violet] (11.1,5) node{UV photons};
    
    \draw[<->, very thick] (0,6) -- (10,6) node[above,midway] {\textbf{Photodissociation region}};
    
    \draw (8.7,4.5) node{H$_\textsc{II}$};
    \draw (2,4.5) node{Molecular Cloud};
    \draw (5.7,4.5) node{Neutral};
    \draw (4.25,5.5) node{DF};
    \draw[>=latex,->] (4.25,5.2) -- (4.25,4.5);
    \draw (7.25,5.5) node{IF};
    \draw[>=latex,->] (7.25,5.2) -- (7.25,4.5);
    
    \draw (8.7,0.8) node{H$^+$};
    \draw (2,0.8) node{H$_2$};
    \draw (5.7,0.8) node{H};
    
    \node[inner sep=0pt] (graph) at (5.3,-4.5)
    {\includegraphics[width=0.61\linewidth]{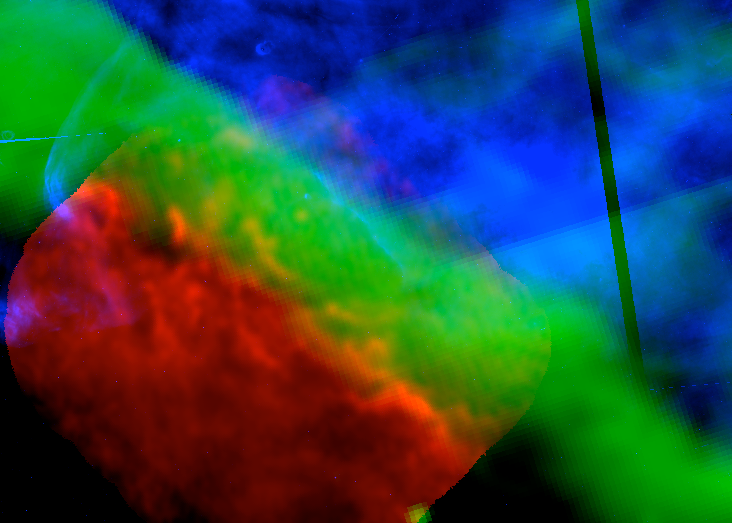}};
    
    
    \draw[orange] (8.7,0.3) node{\Large$\bullet$};
    \draw[orange] (7.25,0.3) node{\Large$\bullet$};
    \draw[orange] (4.25,0.3) node{\Large$\bullet$};
    \draw[orange] (2,0.3) node{\Large$\bullet$};
    
   \draw [dashed,thick,orange] (8.7,0.3) -- (6.8,-1) --  (6.8,-2.77);
    \draw[orange] (6.8,-2.77) node{\Large$\bullet$};
    
    \draw [dashed,thick,orange] (7.25,0.3) -- (5.5,-1) -- (5.5,-4.2);
    \draw[orange] (5.5,-4.2) node{\Large$\bullet$};
    
    \draw [dashed,thick,orange] (4.25,0.3) -- (4.5,-1) -- (4.5,-5.3);
    \draw[orange] (4.51,-5.28) node{\Large$\bullet$};
    \draw [dashed,thick,orange] (2,0.3) -- (2.17,-1) -- (2.17,-7.8);
    \draw[orange] (2.19,-7.8) node{\Large$\bullet$};
    
    \draw [very thick,orange,|-|] (2,-8) -- ++(5.5,6);
    
    \node[inner sep=0pt] (graph) at (5.25,-12.2) 
    {\includegraphics[width=0.61\linewidth]{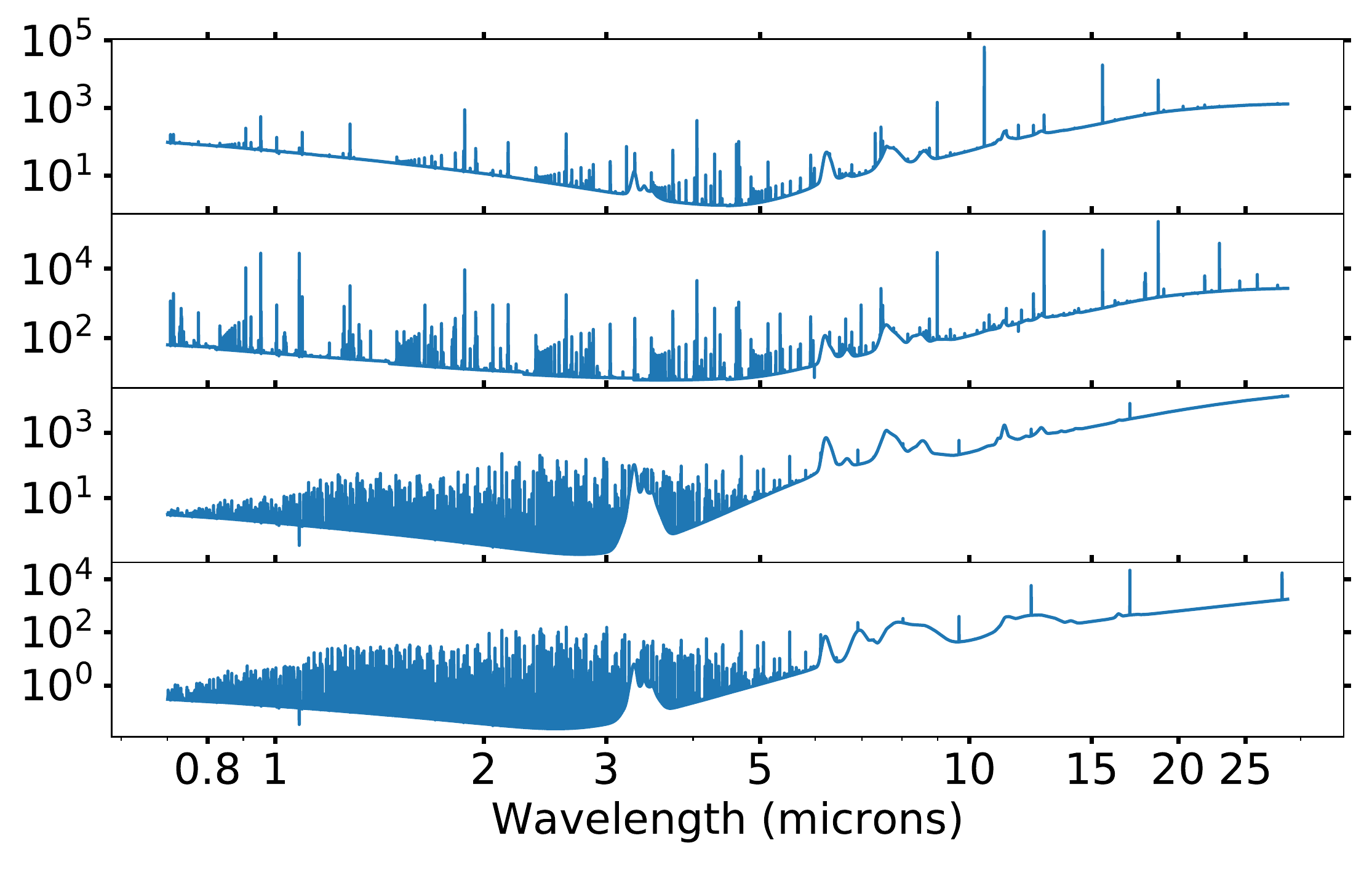}};
    
    \draw[dashed,orange] (6.8,-2.77) -- (6.8,-9.1);
    \draw[orange] (6.8,-9.1) node{\Large$\bullet$};
    
    \draw[dashed,orange] (5.5,-4.2) -- (5.5,-10.6);
    \draw[orange] (5.5,-10.6) node{\Large$\bullet$};
    
    \draw[dashed,orange] (4.5,-5.3) -- (4.5,-12);
    \draw[orange] (4.5,-12) node{\Large$\bullet$};
    
    \draw[dashed,orange] (2.17,-7.8) -- (2.17,-13.6);
    \draw[orange] (2.17,-13.6) node{\Large$\bullet$};
    
    \node[rotate=90,thick] at (-0.4,-11.7) {Intensity (mJy arcsec$^-2$)} ;
    
    \draw[>=latex,->,orange,very thick] (10.51,-8.15) -- ++(0,0.75);
    \draw[>=latex,->,orange,very thick] (10.51,-8.15) -- ++(-0.75,0);
    \node[orange,very thick] at (10.51,-7.2) {N};
    \node[orange,very thick] at (9.6,-8.1) {E};
    
    \draw[->,very thick,rotate=48,orange] (5.1,-7.8) -- ++(0.5,0);
    \node[thick,text width=2.5cm,rotate=48,orange] at (10.1,-1.5) {to Trapezium stars};
    
    \draw[orange] (0,-8.1) -- ++(0.7,0);
    \node[orange] at (0.34,-7.9) {\footnotesize 2 pc};
    
    \end{tikzpicture}
    \caption{Photodissociation Region. \textbf{Top:} Photodissociation region synthetic morphology with, from right to left, ionized or H$_\textsc{II}$ region (H$_\textsc{II}$), ionization front (IF), the neutral region,  dissociation front (DF) and the molecular cloud. \textbf{Middle:} Photodissociation region as seen by HST (in blue) \citep{Bally2015}, the Spitzer Space Telescope \citep{Fazio2004} and ALMA (in red) \citep{Goicoechea2016}. \textbf{Bottom:} 4 synthetic spectra from 0.7 to 28 microns of 4 different regions from the photodissociation region with, from top to bottom, the ionized region, the ionization front, the dissociation front, the molecular cloud (see text for details).}
    \label{pdrfig}
\end{figure*}

In star forming regions, heating processes by extreme UV (EUV, $E<13.6$ eV) and far UV (FUV $E<13.6$ eV) photons give PDRs a specific structure schematically represented in Fig. \ref{pdrfig}. The H$_\textsc{II}$ layer is the region dominated by EUV absorption and is composed of ionized gas. Its temperature is about $10^4$K and its density a few hundred ions per cm$^3$. FUV emissions penetrates deeper in the cloud and heat the neutral region, which is composed of neutral atomic gas. The temperature there is in the range of a few 100 to a few 1000 K and its density between a 1000 and 10$^4$ hydrogen atom per cm$^3$. The limit between these two regions, where protons and electrons recombine, is called the ionization front (denoted IF in  Fig. \ref{pdrfig} ). 
When the amount of FUV photons decreases sufficiently, hydrogen atoms can combine to form dihydrogen molecules (H$_2$). This region is the molecular cloud. The temperature there is between a few 10 to a few 100 K range, and its density is about $10^4$ to $10^6$ molecules per cm$^3$. The limit where hydrogen atoms combine to become dihydrogen molecules, between the neutral region and the molecular cloud, is defined as the dissociation front (denoted by DF in Fig. \ref{pdrfig} ).


\section{\label{synthpdr}Synthesis of a PDR scene: The Orion Bar}
\subsection{Approach}
\renewcommand{\arraystretch}{1.2}
\begin{table*}
    \centering
        \caption{Properties of the synthetic scene of the Orion Bar $\mathbf{X}$ presented in Fig.~\ref{fig:simus} and the underlying elementary spectra and spatial ``weight'' matrices, respectively denoted as $\mathbf{H}$ and $\mathbf{A}$.}
    \begin{tabular}{lccc}
    \hline
    & $\textbf{X}$ & $\textbf{H}$  & $\textbf{A}$  \\
    \hline
    \hline
    Wavelength range ($\mu$m)& 0.7-28.5 & 0.7-28.5 & --\\    
    Spectral Resolution $\left(R=\frac{\lambda}{\Delta\lambda}\right)$& $\sim$ 3000 & $\sim$ 3000 & -- \\
    \hline
    FOV & 31" $\times$ 31" & -- & 31" $\times$ 31" \\
    Pixel size (arcsec$^2$)& 0.031 $\times$ 0.031 & -- & 0.031 $\times$ 0.031\\
    \hline
    Full size (pixels) &  23000 $\times$ 1000 $\times$ 1000  & 23000  $\times$ 4 & 4 $\times$ 1000 $\times$ 1000 \\
    \hline
    \end{tabular}
    \label{tab:recap_scene}
\end{table*}

This section describes the synthesis of an accurate astrophysical scene of infrared emissions of a PDR. Here, we take the canonical PDR of the Orion Bar as a reference for the construction of this synthetic scene.  
This scene consists of a high spatio-spectral resolution 3D cube with 2 spatial dimensions and 1 spectral dimension.
For convenience, the scene is not referred to as a 3D object, but rather as a 2D matrix whose columns contain the spectra associated with each spatial location. More precisely, 
let $\mathbf{X}$ denote the matrix corresponding to the synthetic scene where each column corresponds to the spectrum at a given location. This high spatial and high spectral resolution image is assumed to result from the product
\begin{equation}
    \mathbf{X}=\mathbf{H}\mathbf{A}
\end{equation}
where $\mathbf{H}$ is a matrix of elementary spectra and $\mathbf{A}$ is the matrix of their corresponding spatial ``weight'' maps. The size, spectral range and spatial field of view of these matrices are detailed in Tab.~\ref{tab:recap_scene}. The underlying assumption of this model is that the data follow a linear model, i.e., the spectra composing the scene are linear combinations of spectra coming from ``typical'' regions. This choice has been adopted for several reasons. First, there is no spatio-spectral model of PDRs able to provide computed spectra with all signatures observable at mid-infrared wavelengths (gas lines, PAHs, dust etc.) and accounting for the complex spatial textures generally found in the observations \citep{Goicoechea2016}. The second reason is that the linearity of the mixture is a reasonable assumption at mid-IR wavelengths, where most of the emission is optically thin, except perhaps around $9.7\mu$m where silicate absorption may have an effect for large column densities \citep{Weingartner2001}. An additional advantage of defining $\mathbf{X}$ as a matrix product is related its computing storage: the full matrix is expected to be quite large, and simply impossible to store in memory. Instead, adopting such a decomposition, only the underlying model factors $\mathbf{H}$ and $\mathbf{A}$ need to be stored, hence significantly reducing the occupied memory. The following sections describe the choice of the elementary signatures in $\mathbf{H}$ and their spatial mapping in $\mathbf{A}$.

\subsection{Elementary spectra $\mathbf{H}$}

The elementary spectra composing the matrix $\mathbf{H}$ have been computed within the framework of the ERS 1288 program \citep{Berne2017}. A more detailed description of how they have been calculated will be provided in a paper aiming at describing the scientific objectives of this ERS 1288 program.
This matrix $\mathbf{H}$ is composed of $k=4$ spectra corresponding to four regions of a PDR as depicted in Fig.~\ref{pdrfig}: the H$_\textsc{II}$ region, the ionization front, the dissociation front and the molecular cloud. These spectra have been computed individually for each region, using the Meudon PDR code for the contribution from molecular and atomic lines \citep{Lepetit2006}, the CLOUDY code for the ionized gas \citep{Ferland1998}, the PAHTAT model for the PAH emission \citep{Pilleri2012}, and the DUSTEM model for the contribution from the dust continuum \citep{Compiegne2011}. The physical parameters used for these models correspond to those of the Orion Bar, which is a well studied region. Absolute calibration of the resulting spectra depicted in Fig.~\ref{pdrfig} (bottom) for each one of the four regions has been crossed-check with existing observations of the Orion Bar, so as to confirm that they are realistic in terms of flux units.

\subsection{Abundance Maps ($\mathbf{A}$)}

Since the spectra in $\mathbf{H}$ carry the flux information, spatial abundance maps in the matrix $\mathbf{A}$ correspond to normalized between 0 and 1 textures. In this work they are derived from real data from the Hubble Space Telescope (HST) and the Atacama Large sub-Millimeter Array (ALMA). For clarity, they have been roated to obtain a plane-parallel morphology reminiscent of the conceptual structure described in Fig.~\ref{pdrfig}. This is also because the FOV of the currently planned observation of the ERS 1288 program will be perpendicular to the IF/DF, i.e., corresponding to a horizontal cut in the rotated images.  The chosen FOV for the synthesis of the texture maps from the observations is a 30 $\times$ 30 arcsec$^2$, square centered on coordinates $\mathrm{RA}=5:35:20.0774$  $\mathrm{DEC}=-5:25:13.785$ in Orion. The textures associated with the four spectral components are described below according to their corresponding regions.

\subsubsection{Ionization front \& H$_\textsc{II}$ region}

\begin{figure*}
    \centering
    \begin{tikzpicture}
    \node[inner sep=0pt] (graph) at (-2,0)
    {\includegraphics[width=0.45\linewidth]{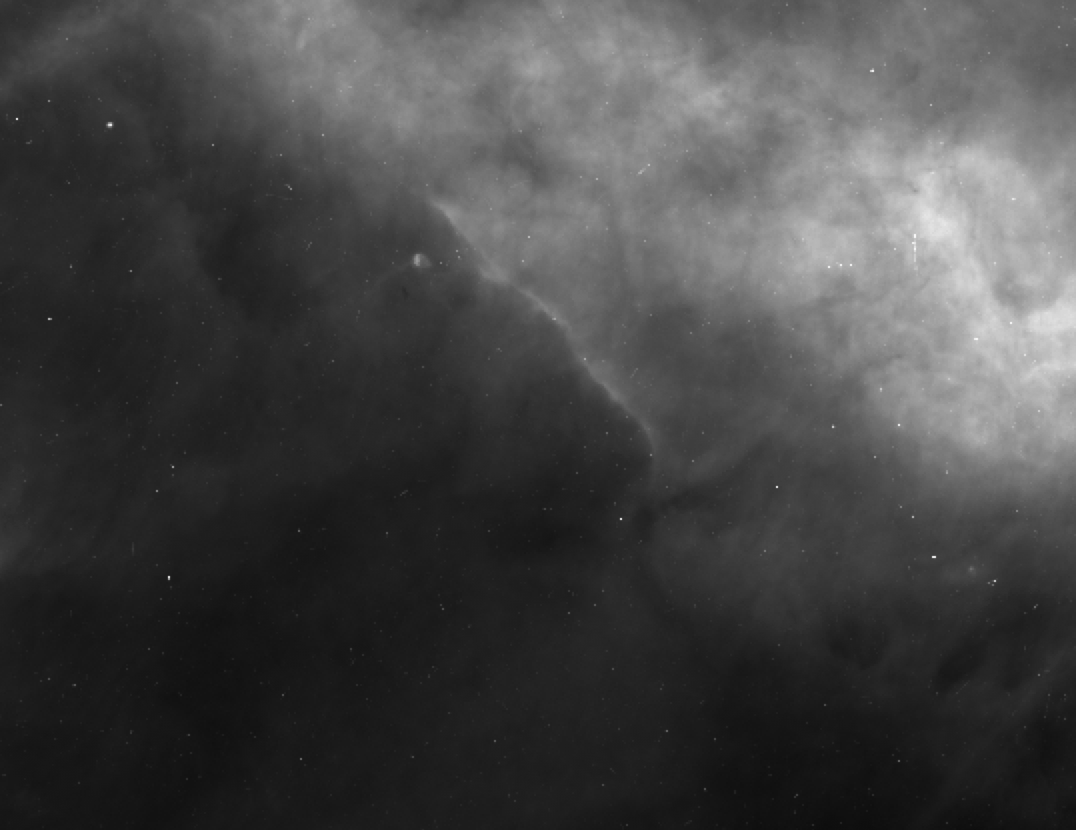}};
    \node[inner sep=0pt] (graph) at (8,2.8)
   {\includegraphics[width=0.3\linewidth]{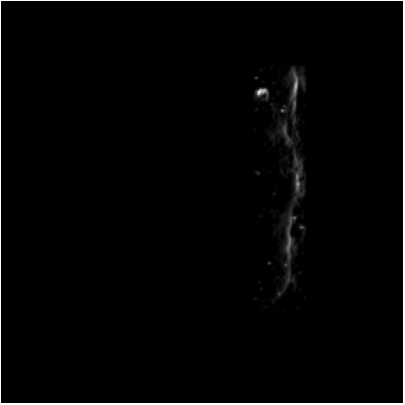}}; 
    \node[inner sep=0pt] (graph) at (8,-2.8)
   {\includegraphics[width=0.3\linewidth]{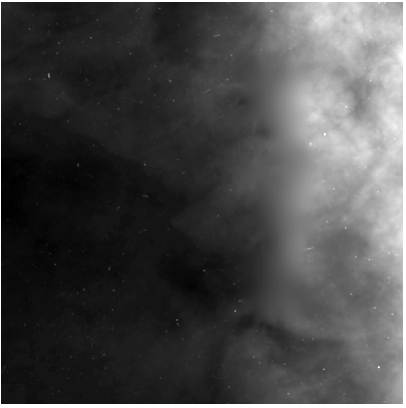}}; 
   \node[thick] at (-6.4,2.9) {a)};
   \node[thick] at (5,5.3) {b)};
   \node[thick] at (5,-0.3) {c)};
   \draw[->,thick] (2.5,0) -- ++(2,2);
   \draw[->,thick] (2.5,0) -- ++(2,-2);
    \node[rotate=0,thick,orange] at (7,5.3) {Ionization front texture} ;
    \node[rotate=0,thick,orange] at (6.6,-5.2) {H$_\textsc{II}$ Region texture} ;
    \node[rotate=0,thick,orange] at (-5,2.9) {HST image} ;
    \draw[->,very thick,rotate=48,orange] (2.5,1.1) -- ++(0.5,0);
    \node[thick,text width=2.5cm,rotate=48,orange] at (1.6,2.5) {to Trapezium stars};
    \draw[very thick,orange,rotate=41] (-4.1,-0.5) rectangle ++(4.4,4.4);

    \draw[>=latex,->,orange,thick] (1.95,-3) -- ++(0,0.75);
    \draw[>=latex,->,orange,thick] (1.95,-3) -- ++(-0.75,0);
    \node[orange,thick] at (1.85,-2.1) {N};
    \node[orange,thick] at (1.05,-3) {E};
    \end{tikzpicture}
    \caption{\textbf{a)}: HST image of 656nm H$_\mathrm{\alpha}$ emission line in the Orion Bar \citep{Bally2015} and the chosen FOV (orange box). \textbf{b)}: Extracted ionization front texture. \textbf{c)}: Extracted H$_\textsc{II}$ region texture. Normalized scale (black: 0; white: 1), centered in RA=5:35:20.0774 DEC=-5:25:13.785}
    \label{hst_front}
\end{figure*}

To build an accurate spatial representation of the H$_\textsc{II}$ region and the ionization front, we have resorted to the Orion Bar image obtained by the narrow band filter centered at $656$nm (H$_\mathrm{\alpha}$ emission line) of the WFC3 instrument aboard the HST (Fig. \ref{hst_front}). This image was taken as part of the observing proposal lead by \cite{Bally2015}.
This image provides an accurate view of the morphology of the H$_\textsc{II}$ region and the ionization front combined \citep{Tielens2005}. 

After cropping and rotating, we have separated the ionization front and H$_\textsc{II}$ region in the observed image. 
As the brightness of the ionization front is comparable to the brightest regions in the H$_\textsc{II}$ region, a thresholding on the raw image does not isolate efficiently the ionization front from the plasma cloud. However, unlike the H$_\textsc{II}$ region, the front appears as a sharp line where the gradient of the image is high. Therefore, the location of the pixels in the image belonging to the IF can be easily recovered from the pixel-scale horizontal gradient of the image. Thus, the latter is thresholded to create a mask around the pixels with highest gradient magnitudes. The smallest connected components (smaller than 10000 pixels) are then removed to delete high gradient values related to small objects in the image and thus not related to the IF. Finally, the original HST image is term-wise multiplied by this mask and slightly smoothed by a Gaussian kernel with full width at half-maximum (FWHM) of 4.7 pixels.
This allows the contribution  to be extracted from the IF only, and finally an IF image to be obtained (see Fig.~\ref{hst_front}).  

Once the ionization front is removed from the original image, the gap is expended by a morphological dilation with a $20$ pixels-diameter disk and filled using a standard inpainting technique \citep{Damelin2017}.
This process fills the missing part by selecting similar textures available outside the mask. The result is shown in Fig \ref{hst_front}. Both images are then up-sampled to the resolution of the JWST NIRCam Imager instrument by bi-cubic spline interpolation with an 1.25 up-sampling factor. 

\subsubsection{Dissociation front}

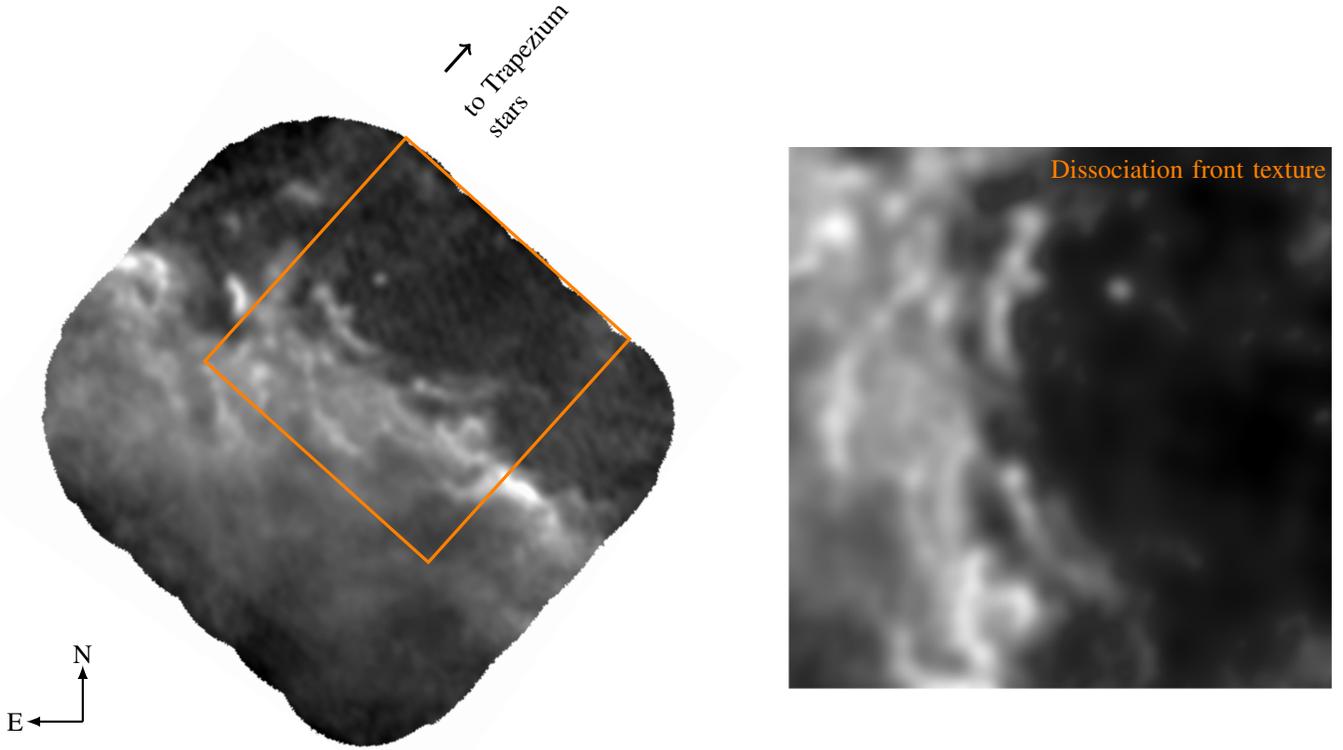
\begin{figure*}
    \centering
    \hspace{-1.1cm}
    \begin{tikzpicture}
    \node[inner sep=0pt] (graph) at (-0.5,0)
    {\includegraphics[width=0.4\linewidth,angle=55]{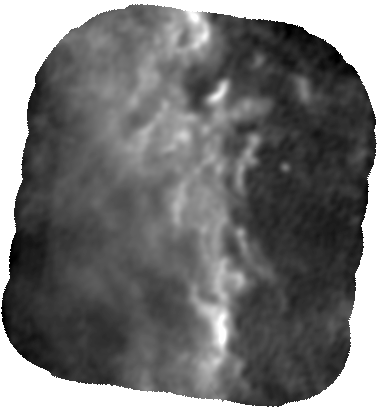}};
    \node[inner sep=0pt] (graph) at (9,0)
    {\includegraphics[width=0.4\linewidth,angle=0]{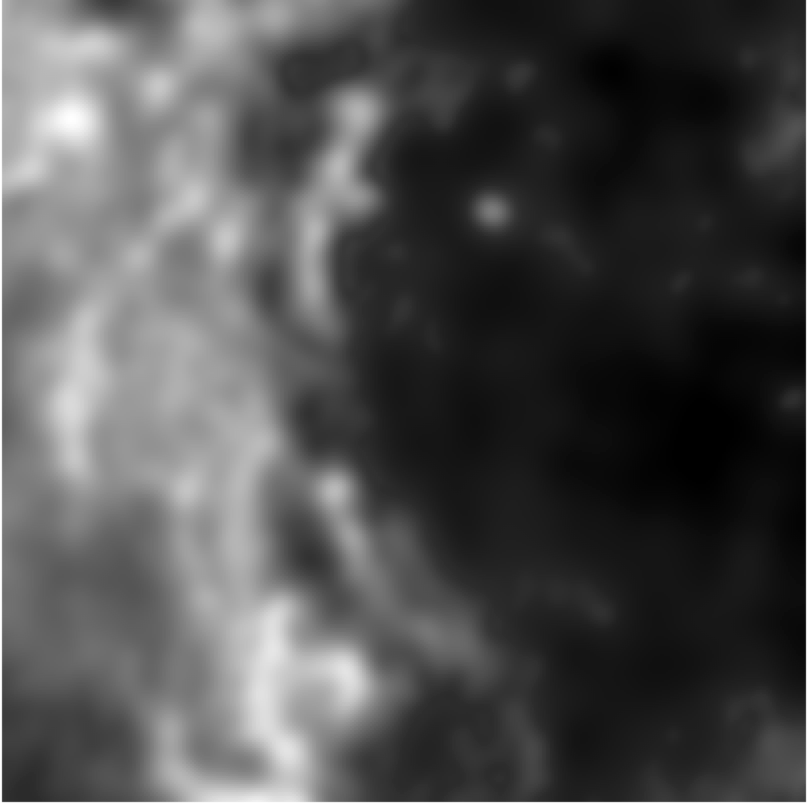}};
    
    \draw[very thick,orange,rotate=48] (-1,-1.7) rectangle ++(4,4);
    
    \draw[->,very thick,rotate=48] (4,2.5) -- ++(0.5,0);
    \node[thick,text width=2.5cm,rotate=48] at (2.1,4.9) {to Trapezium stars};
    
    \draw[>=latex,->,thick,rotate=0] (-4,-4) -- ++(0,0.75);
    \draw[>=latex,->,thick,rotate=0] (-4,-4) -- ++(-0.75,0);
    \node[thick,rotate=0] at (-4,-3.1) {N};
    \node[thick,rotate=0] at (-4.9,-4) {E};
    
    \node[thick,orange] at (10.7,3.35) {Dissociation front texture};
    
    \end{tikzpicture}
    \caption{\textbf{Left}: ALMA image of HCO 3-2$^+$ line peak in the Orion Bar \citep{Goicoechea2016} and the chosen FOV (orange box). \textbf{Right}: Normalized texture for the dissociation front abundance map (black=0, white=1).}
    \label{alma_df}
\end{figure*}

The texture map related to the dissociation front is derived from an image of HCO (3-2)$^+$ emission observed  by \citet{Goicoechea2016} with ALMA, see Fig. \ref{alma_df}. According to the authors, this map locates well the H/H$_2$ transition and is consequently used here to define the dissociation front of this PDR. 

After rotation and cropping, the high textured zone in the right part of the chosen area is extracted by thresholding. Then it is slightly smoothed by a Gaussian kernel with a 2.3 pixels FWHM. The remaining part, less structured, is much more smoothed thanks to a Gaussian kernel with a 9.4 pixels FWHM to remove visible noisy stripes due to ALMA data acquisition process. 
 
Then, the smoothed image is up-sampled to the resolution of the NIRCam Imager instrument by bi-cubic spline interpolation with an up-sampling factor of 5 and normalized. The resulting texture is shown Fig. \ref{alma_df} (right).

\subsubsection{Molecular cloud}

\begin{figure*}
    \centering
    \hspace{-1.1cm}
    \begin{tikzpicture}
    \node[inner sep=0pt] (graph) at (-0.5,0)
    {\includegraphics[width=0.48\linewidth,angle=2]{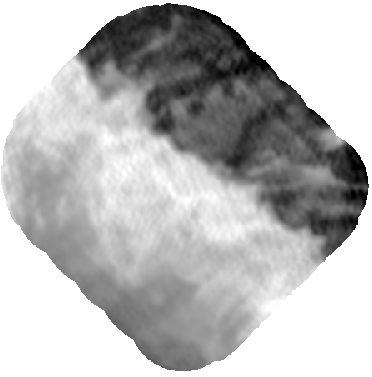}};
    \node[inner sep=0pt] (graph) at (9,0)
    {\includegraphics[width=0.4\linewidth]{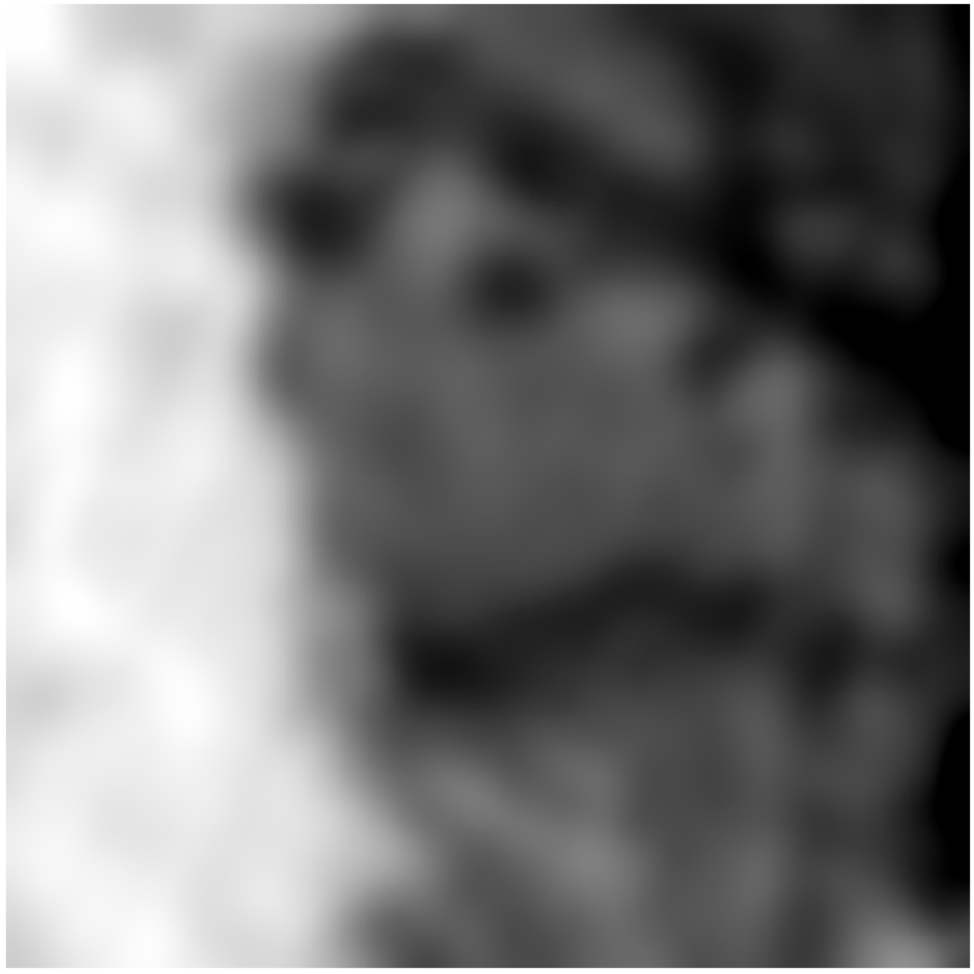}};
    
    \draw[very thick,orange,rotate=48] (-0.9,-1.3) rectangle ++(4,4);
    
    \draw[->,very thick,rotate=48] (4,2.5) -- ++(0.5,0);
    \node[thick,text width=2.5cm,rotate=48] at (2.1,4.9) {to Trapezium stars};
    
    \draw[>=latex,->,thick,rotate=0] (-4,-4) -- ++(0,0.75);
    \draw[>=latex,->,thick,rotate=0] (-4,-4) -- ++(-0.75,0);
    \node[thick,rotate=0] at (-4,-3.1) {N};
    \node[thick,rotate=0] at (-4.9,-4) {E};
    
    \node[thick,orange] at (10.7,3.35) {Molecular cloud texture};

    \end{tikzpicture}
    \caption{\textbf{Left}: ALMA image of CO 3-2 line peak in the Orion Bar \citep{Goicoechea2016} and the chosen FOV (orange box). \textbf{Right}: Normalized texture molecular cloud abundance map (black=0, white=1).}
    \label{alma_mc}
\end{figure*}
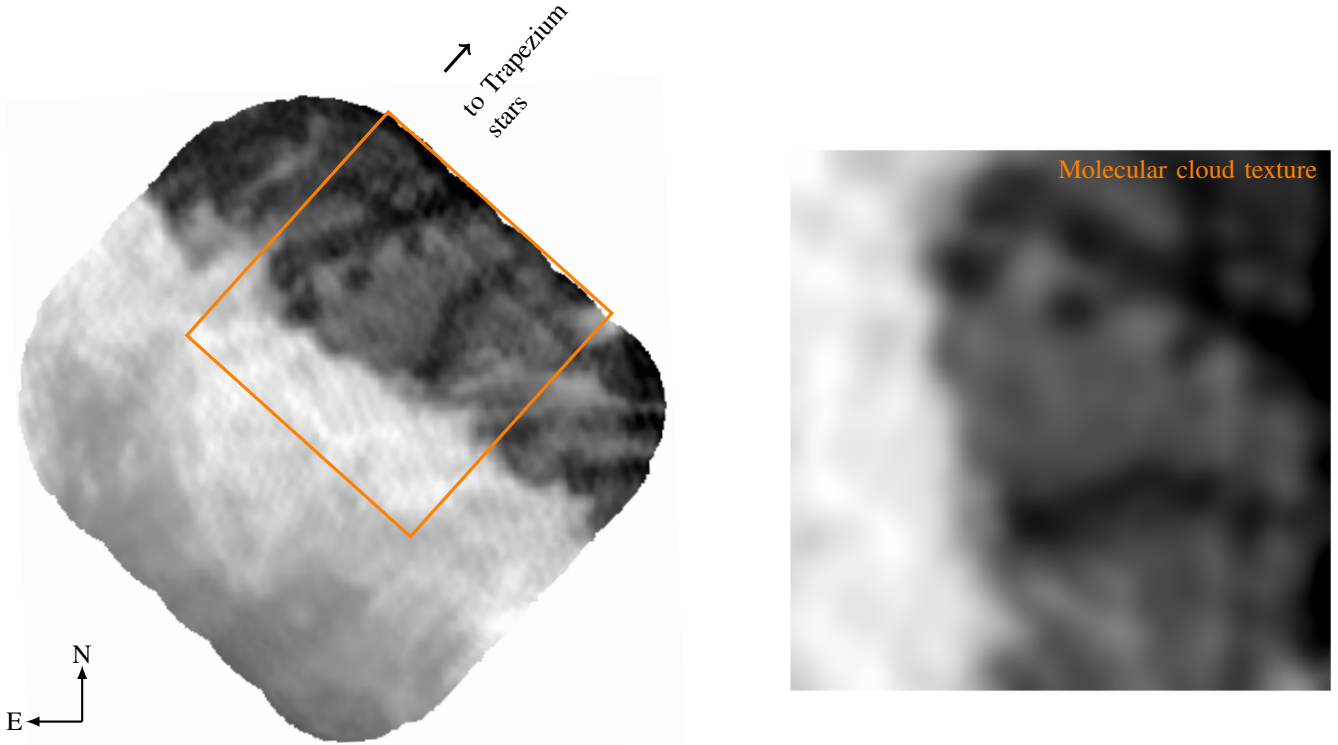

The molecular cloud texture map has been also extracted from an ALMA image \citep{Goicoechea2016}. 
The CO (3-2) emission line is commonly used as a tracer of the molecular cloud.
As explained in the previous section, the area in the orange box in Fig. \ref{alma_mc} (left) has been chosen to match the textures maps already build. The stripes due to ALMA data acquisition process are clearly noticeable over the full FOV. Therefore, after rotation, the image is strongly smoothed with a Gaussian kernel with a 11 pixels FWHM to remove these unwanted stripes, identified as noise. The smoothed image is finally up-sampled to the resolution of the NIRCam Imager instrument by bi-cubic spline interpolation (with an up-sampling factor of 5) and normalized. The resulting texture is shown Fig. \ref{alma_mc} (right).  

\section{\label{formod}JWST instruments forward model}
\label{sec:forward}
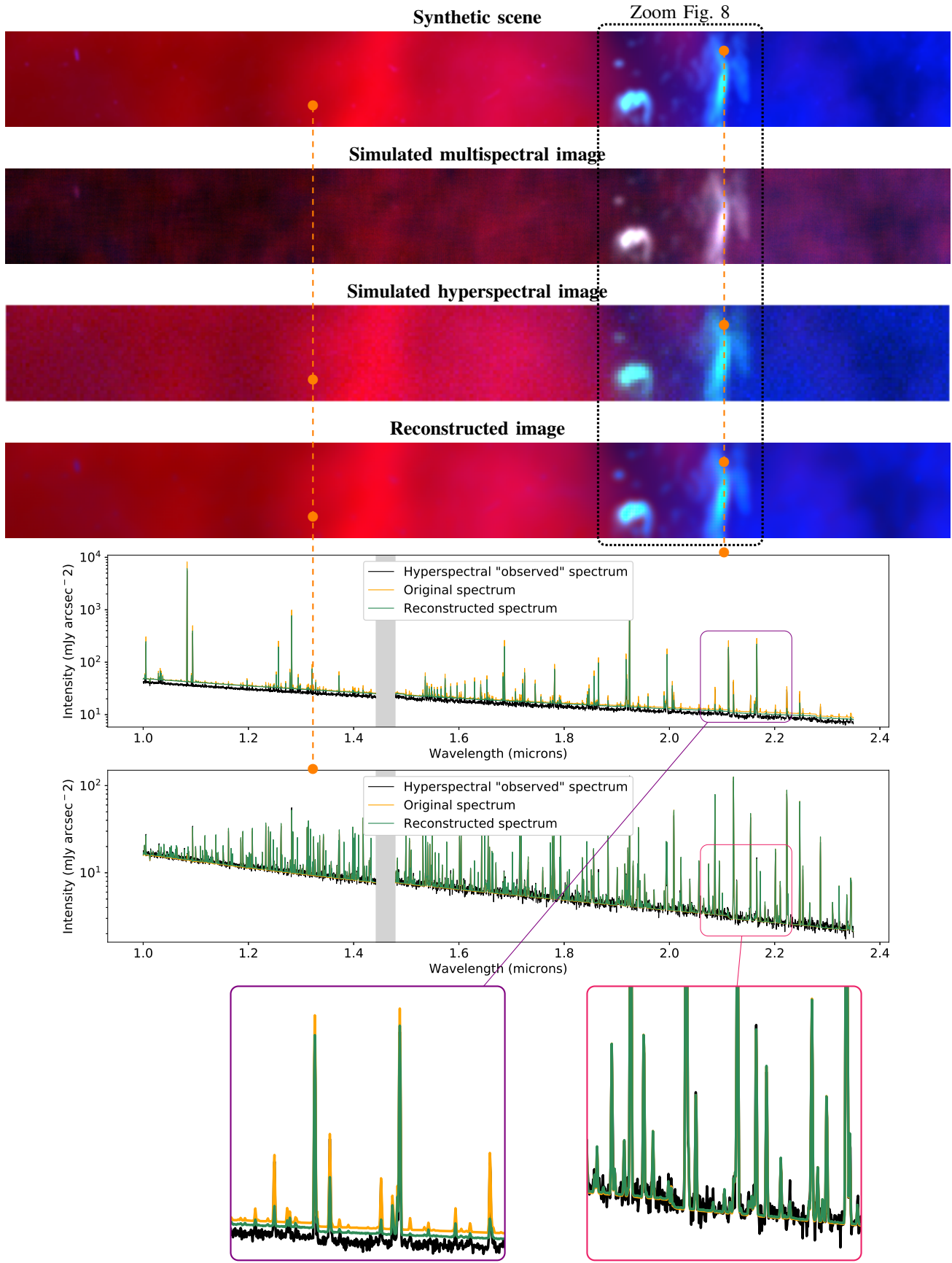
\begin{figure*}
    \centering
    \begin{tikzpicture}[spy using outlines={size=5cm, connect spies, rounded corners, thick}]
    
    \node[inner sep=0pt] (graph) at (0,2.5) {\includegraphics[width=0.95\linewidth]{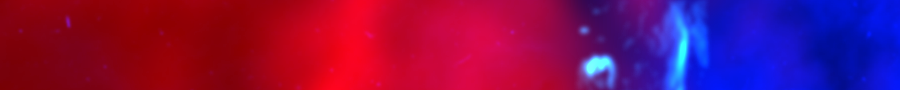}};
    \node[inner sep=0pt] (graph) at (0,0) {\includegraphics[width=0.95\linewidth]{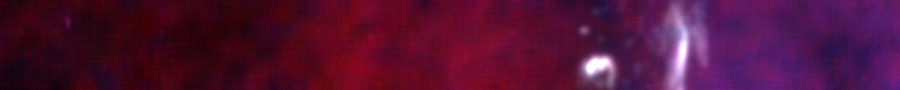}};
    \node[inner sep=0pt] (graph) at (0,-2.5) {\includegraphics[width=0.95\linewidth]{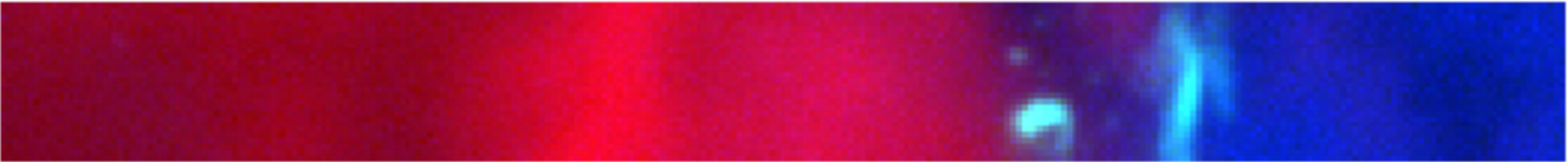}};
    \node[inner sep=0pt] (graph) at (0,-5) {\includegraphics[width=0.95\linewidth]{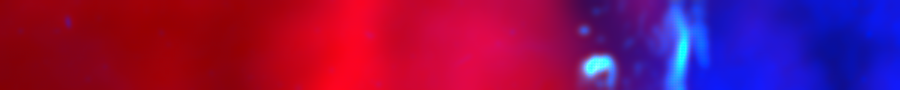}};
    
    \node[thick] at (0,3.6) {\textbf{Synthetic scene}};
    \node[thick] at (0,1.1) {\textbf{Simulated multispectral image}};
    \node[thick] at (0,-1.4) {\textbf{Simulated hyperspectral image}};
    \node[thick] at (0,-3.9) {\textbf{Reconstructed image}};

    \node[inner sep=0pt] (graph) at (0,-10) {\includegraphics[width=0.85\linewidth]{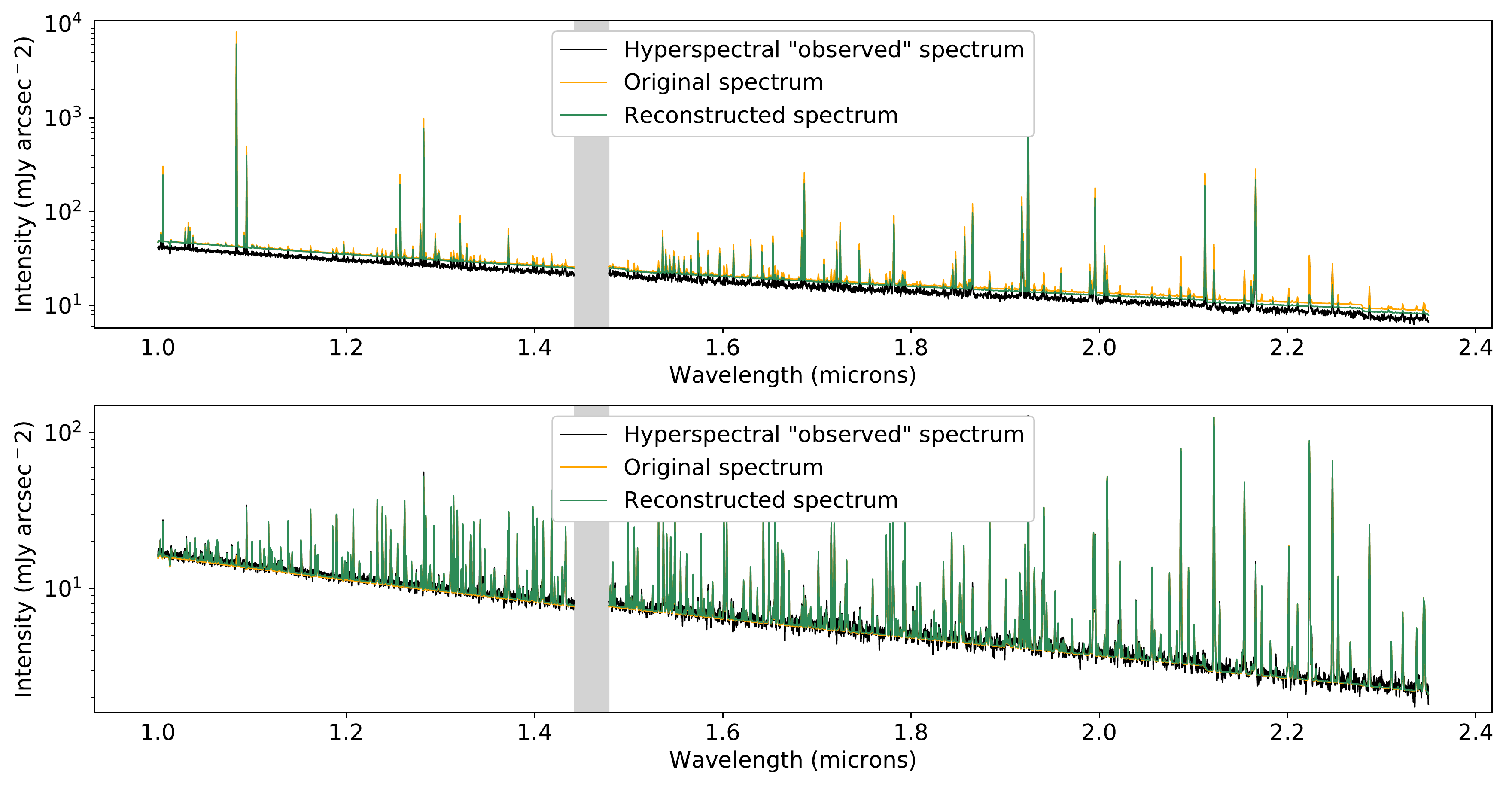}};
    
    \draw[densely dotted, very thick, rounded corners, black] (2.2,3.5) rectangle (5.2,-6);
    \node[black] at (3.7,3.7) {Zoom Fig. \ref{zoom}};
    
    \draw[dashed,orange,thick] (4.5,3) -- (4.5,-6.15);
    \node[orange] at (4.5,3) {\Large$\bullet$};
    \node[orange] at (4.5,3-5) {\Large$\bullet$};
    \node[orange] at (4.5,3-7.5) {\Large$\bullet$};
    \node[orange] at (4.5,-6.15) {\Large$\bullet$};
    
    \draw[dashed,orange,thick] (-3,2) -- (-3,-10.1);
    \node[orange] at (-3,2) {\Large$\bullet$};
    \node[orange] at (-3,2-5) {\Large$\bullet$};
    \node[orange] at (-3,2-7.5) {\Large$\bullet$};
    \node[orange] at (-3,-10.1) {\Large$\bullet$};
    
    \spy[violet, magnification=3] on (4.9,-8.4) in node at (-2,-16.55);
    \spy[WildStrawberry, magnification=3] on (4.9,-12.3) in node at (4.5,-16.55);

    \end{tikzpicture}
    \caption{\textbf{Top} (from top to bottom)\textbf{:} RGB compositions of the synthetic PDR scene, the simulated NIRCam Imager multispectral image, the simulated NIRSpec IFU hyperspectral image and the fused image of high spatio-spectral resolution. Red: H$_2$ emission line pic intensity at 2.122 $\mu$m, Green: H recombination line pic intensity at 1.865 $\mu$m, Blue: Fe$^+$ emission line pic intensity at 1.644 $\mu$m. \textbf{Bottom: } Calibrated spectra associated with the simulated hyperspectral image, original and reconstructed spectra related to two pixels of the images above. Zoom parts are 3 times magnified. Intensities are plotted in a logarithmic scale.}
    \label{fus_res}
\end{figure*}

In this section, we derive a simple yet sound mathematical model of two instruments embedded in the JWST, namely NIRSpec IFU and NIRCam Imager. A more advanced modeling was previously conducted by the teams in charge of the JWST Exposure Time Calculator (ETC)  at the space telescope science institute (STScI) \citep{Pontoppidan2016}. The ETC is a tool for astronomers to simulate data acquisition and to compute signal-to-noise ratios (SNR) for all JWST observing modes and instruments. This tool models the full acquisition process (groups, integration ramps) and noise for astrophysical scenes composed of complex spectra and several extended (ellipses) or point sources. However, the ETC tool exhibits two major limitations in the context of the work targeted in this paper, i.e., within a fusion perspective. First, currently there is no stable version of the ETC that provides simulated measurements for complex spatio-spectral 3D scenes such as the astrophysical scene described in Sect.~\ref{synthpdr}. Note that we are currently working with STScI to overcome this limitation, e.g., by using the linear properties of the synthetic scene described in section \ref{synthpdr}. The second reason is that the forward models involved in the considered fusion method requires to be explicit hence less advanced than those provided by the ETC (see Sect. \ref{fus_res}). As a consequence, we derived explicit forward models capitalizing on the information available in the ETC as a reference. The models associated with the two instruments under consideration, supplemented by a suitable noise modeling, are described in what follows. 


\begin{table*}
    \centering
        \caption{Main technical features of NIRCam Imager and NIRSpec IFU considered in this study.}
    \begin{tabular}{lccc}
    \hline
    & \multicolumn{2}{c}{\textbf{NIRCam Imager}} & \textbf{NIRSpec IFU} \\
    \hline
    \hline
    Channel & SW & LW & \\
    \hline
    Wavelength range ($\mu$m)& 0.6-2.35 & 2.35-5 & 0.6-5.3\\
    Spectral Resolution $\left(R=\frac{\lambda}{\Delta\lambda}\right)$& $\sim$ 1-100 & $\sim$ 1-100 & $\sim$ 3000 \\
    Spectral points & 13 & 16 & $\sim$ 12000 \\
    \hline
    FOV & 2.2' $\times$ 5.1' (with gaps) & 2.2' $\times$ 5.1' (with gaps)& 3" $\times$ 3" \\
    Pixel size (arcsec$^2$)& 0.031 $\times$ 0.031 & 0.063 $\times$ 0.063 & 0.1 $\times$ 0.1\\
    FOV (pixels) & 8 $\times$ 2040 $\times$ 2040 & 2 $\times$ 2040 $\times$ 2040 & 30 $\times$ 30 \\
    \hline
    \end{tabular}
    \label{recap_nir}
\end{table*}

\subsection{NIRCam Imager}

The near-infrared camera NIRCam Imager aboard the JWST will observe space from 0.6 to 5 microns with 3 possible data acquisition modes: imaging, coronagraphy and slit-less spectroscopy. The observing mode studied in this paper is the imaging mode. 
It will provide multispectral images  on wide fields of view (2.2' $\times$ 5.1', separated on two adjacent modules). This instrument covers the 0.6 to 5 microns wavelength range simultaneously through 2 channels, the SW channel between 0.6 and 2.3 microns and the long wavelength channel (LW) between 2.4 and 5 microns, via $l_\mathrm{m} = $ 29 extra-wide, wide, medium and narrow filters. Each channel, SW or LW, acquire images composed of $p_m$ pixels with pixel sizes of  0.031~$\times$~0.031 arcsec$^2$ and 0.063~$\times$~0.063 arcsec$^2$, respectively. The main technical features are summarized in Table \ref{recap_nir}.

The proposed mathematical model of the NIRCam Imager detailed in this section is derived to reflect the actual light path through the telescope and the instrument and the corresponding spatial and spectral distortions. The optical system of the telescope and the instrument, and more specifically mirrors, disturb the incoming light and its path. The effect on the detector and therefore on the observed image is a spatial spread of the light arising from the sky, resulting in a blurring of the spatial details. This blurring depends on the wavelength $\lambda$ (in meters) of the incoming light and the JWST primary mirror diameter $D$ (in meters) such that the effective angular resolution $\theta$ (in radians), i.e. the ability to separate two adjacent points of an object, is limited by diffraction. After the optics and the mirrors, the light passes through band-pass filters defined by specific wavelength ranges. 

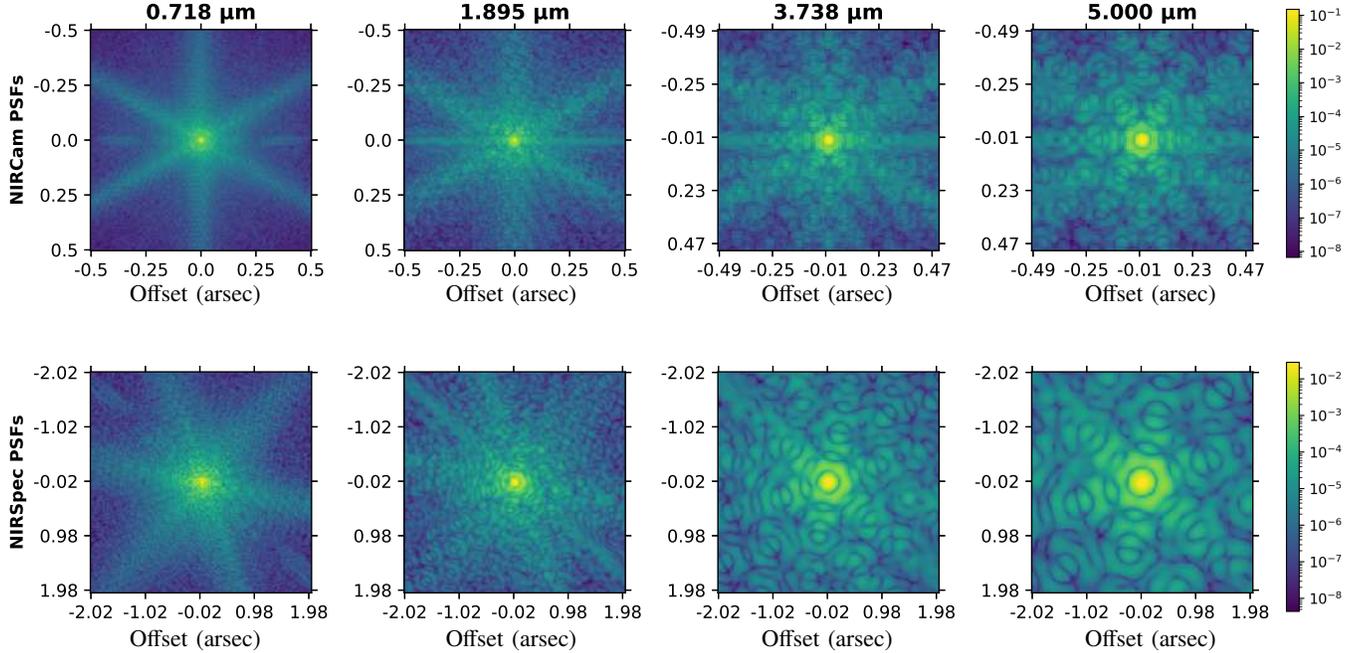
\begin{figure*}
    \centering
    \begin{tikzpicture}
    \node[inner sep=0pt] (graph) at (0,0)
    {\includegraphics[width=0.95\linewidth]{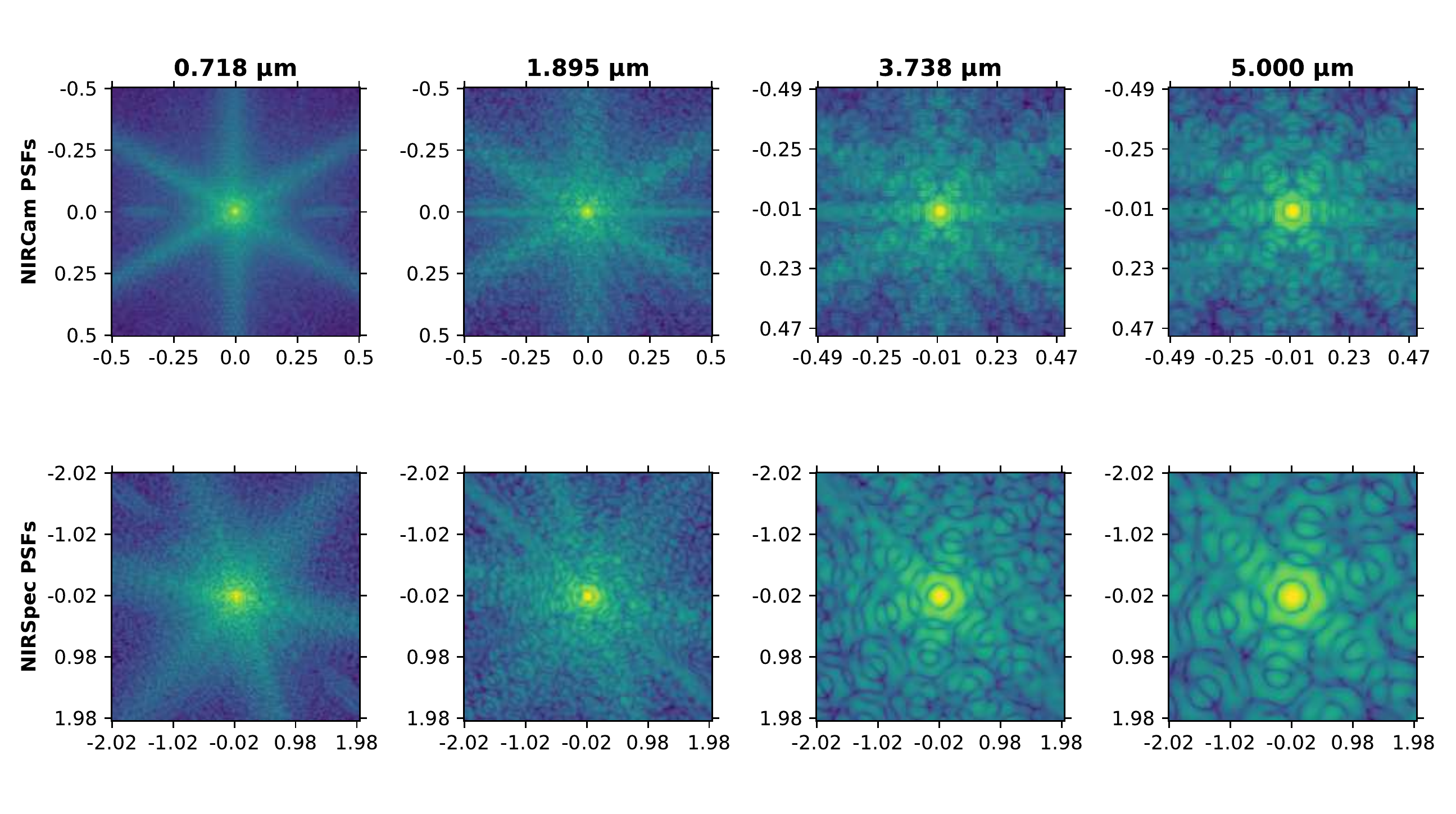}};
    
    \node[] at (-5.9,0.25) {\small Offset (arsec)};
    \node[] at (-5.9,-4.35) {\small Offset (arsec)};

    \node[] at (-1.7,0.25) {\small Offset (arsec)};
    \node[] at (-1.7,-4.35) {\small Offset (arsec)};
    
    \node[] at (2.5,0.25) {\small Offset (arsec)};
    \node[] at (2.5,-4.35) {\small Offset (arsec)};
    
    \node[] at (6.8,0.25) {\small Offset (arsec)};
    \node[] at (6.8,-4.35) {\small Offset (arsec)};
    
    \node[inner sep=0pt] (graph) at (9,2.4)
    {\includegraphics[width=0.048\linewidth]{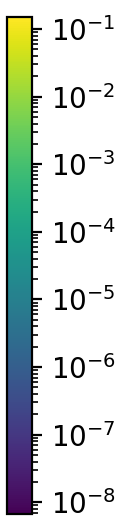}};
    
    \node[inner sep=0pt] (graph) at (9,-2.3)
    {\includegraphics[width=0.045\linewidth]{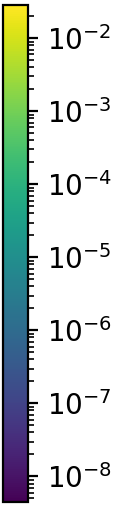}};

    \end{tikzpicture}
    \caption{NIRCam Imager ($1$st row) and NIRSpec IFU ($2$nd row) PSFs calculated with \emph{webbpsf} \citep{Perrin2012} from 0.718 to 5.000 microns (logarithmic scale).}
    \label{psfs}
\end{figure*}

These two main degradations (i.e., spatial blurring and spectral filtering) can be expressed with closed-form mathematical operations successively applied to the astrophysical scene $\mathbf{X}$. 
First, the light spread effect due to the optical systems is modeled as a set of spectrally-varying 2-D spatial convolutions, denoted $\mathcal{M}(\cdot)$. The corresponding PSFs, calculated with the online tool {\it{webbpsf}} \citep{Perrin2012}, are wavelength-dependent and the FWHM of the spread patch grows linearly with the wavelength. This dependency is illustrated in Fig. \ref{psfs} (top) which exhibits the significantly different patterns of four PSF associated with four particular wavelengths. The following spectral filtering step, which degrades the spectral resolution of the scene, can be modelled as multiplications by the transmission functions of the NIRCam Imager filters \citep{JDoxNIRCamFilters}. This operation can be formulated through a product by the matrix $\mathbf{L}_\mathrm{m}$, whose rows are defined by these transmission functions. Therefore, the noise-free multispectral image $\bar{\mathbf{Y}}_\textrm{m}$ composed of $l_\mathrm{m}$ ($\ll l_\mathrm{h}$) spectral bands and $p_\mathrm{m}$ pixels can be written as
\begin{equation}
    \bar{\mathbf{Y}}_\textrm{m} = \mathbf{L}_\textrm{m}\mathcal{M}(\mathbf{X})
\end{equation}
Note that a similar approach was followed by \cite{HadjYoucef2018} to derive the forward model associated with the imager embedded in MIRI. 

\subsection{NIRSpec IFU}

The near-infrared spectrograph NIRSpec IFU embedded in the JWST will perform spectroscopy from 0.6 to 5.3 microns at high (R$\sim$3000), medium (R$\sim$1000) or low (R$\sim$100) spectral resolution
through 4 observing modes. The ERS 1288 program proposed by \cite{Berne2017} will rely on the integral field unit (IFU) with high resolution configuration. It will provide spectroscopic (also called hyperspectral) images on small fields of view (3 $\times$ 3 arcsec$^2$). Data acquisition on the wavelength range is covered by several disperser-filter combinations with similar features. Although unprecedented, the spatial sampling of the IFU is about 9 times less than NIRCam Imager with a 0.1 $\times$ 0.1 arcsec$^2$ pixel size. The main technical features of NIRSpec IFU are summarized and compared with NIRCam Imager features in Table \ref{recap_nir}.

As for NIRCam Imager, the light path from the observed scene to the detector through the telescope and the instrument can be formulated thanks to simple mathematical operations while preserving physical accuracy of the model. At first, the light is spread by the optical system, depending on its wavelength and JWST primary mirror. Secondly, the path through the disperser-filter pair attenuates the light. Finally, this light comes on the detector, which can provide $\sim$ 12000 spectra over a 30 $\times$ 30 pixels$^2$ spatial area.

The optical system distortion effect of the telescope and the instrument on the light is, as  for NIRCam Imager, modeled as a set of 2-D spatial wavelength-dependent convolutions denoted $\mathcal{H}(\cdot)$ with PSFs illustrated in Fig. \ref{psfs} (bottom). The light attenuation induced by the disperser-filter pair is a matrix multiplication by $\mathbf{L}_\mathrm{h}$, whose diagonal is the combination of both transmission functions \citep{JDoxNIRSpecFilters}. Besides attenuation, the physical gaps between NIRSpec IFU detectors involve holes in spectra. These holes are modeled by a null transmission at the corresponding wavelengths.
The spatial response of the detector is seen as a downsampling operator $\mathbf{S}$, which keeps one pixel over a 0.1 $\times$ 0.1 arcsec$^2$ area, after averaging pixels over this area. Finally, the noise-free hyperspectral image $\bar{\mathbf{Y}}_\textrm{h}$ composed of $l_\mathrm{h}$ spectral bands and $p_\mathrm{h}$ $\ll p_\mathrm{m}$  pixels can be written as
\begin{equation}
    \bar{\mathbf{Y}}_\textrm{h}=\mathbf{L}_\textrm{h}\mathcal{H}(\mathbf{X})\mathbf{S}.
    \label{hs_eq}
\end{equation}

\subsection{Noise modeling}

This section models the noise associated with NIRCam Imager and NIRSpec IFU which corrupts the noise-free images $\bar{\mathbf{Y}}_\textrm{m}$ and $\bar{\mathbf{Y}}_\textrm{h}$ to yield the simulated images ${\mathbf{Y}}_\textrm{m}$ and ${\mathbf{Y}}_\textrm{h}$, respectively. This composite model relies on the most commonly used hypotheses on the nature of the space observation noise and on more specific assumptions regarding the JWST detectors. The proposed model neglects some other sources of noise which are more difficult to characterize, e.g., related to cosmic rays and background. A more realistic and exhaustive noise modelling is provided by the STScI via the ETC \citep{Pontoppidan2016}. 

\subsubsection{Quantum noise}

Since the detectors are photon counting devices, the particular nature of light emission conventionally induces observations that obey a Poisson distribution $\mathcal{P}(\bar{y})$ 
whose mean is equal to the photon count $\bar{y}$. In a high flux regime, i.e., when the photon count $\bar{y}$ is typically higher than $20$, this Poisson process can be approximated by an additive heteroscedastic Gaussian noise $\mathcal{N}(\bar{y},\bar{y})$ whose mean and variance is the photon count $\bar{y}$. In the particular context of this work, we will assume that the observations follow this high flux regime, which is a reasonable assumption especially for a very bright source such as the Orion Bar. As a consequence, in practice, the incoming flux $\bar{y}$ in a given pixel and a given spectral band will be corrupted by a random variable drawn from $\mathcal{N}(\bar{y}, \bar{y})$. This model is commonly used to define noise in astronomical imaging \citep{Starck2006}.

\subsubsection{Readout noise}

The main source of corruptions induced by the detectors is a readout noise, which is modeled as an additive, centered, colored Gaussian noise. The correlation between two measurements at given spatial and spectral locations of the observed multiband image can be accurately characterized after unfolding the 3D data cube onto the detector plan. Indeed, the JWST and the ETC documentations \citep{Pontoppidan2016} provide a set of matrices reflecting the expected correlations between measurements at specific positions in the plan of the detectors associated to NIRCam Imager and NIRSpec IFU. These correlation matrices are functions of intrinsic characteristics of the readout pattern, such as the integration time, the number of frames and the number of groups \citep{Rauscher2007}. For a given experimental acquisition setup, the covariance matrix of the additive Gaussian readout noise could be computed after a straightforward ordering of these correlations with respect to the reciprocal folding procedure. Alternatively, this colored Gaussian noise can be added to the unfolded multiband images with a covariance matrix directly defined by the correlations expressed in the detector plan and specified by \cite{Pontoppidan2016}. Complementary information regarding the NIRCam Imager and NIRSpec IFU readout noises is given in what follows.



\paragraph{NIRCam Imager readout noise} --  
The spectral bands of the multispectral image are acquired successively such that the incident image on the detector corresponds to a 2D spatial image in a given spectral. Hence, the induced readout noise is only spatially correlated and can be generated for each spectral band independently. Finally, the covariance matrix describing the spatial correlation of the additive Gaussian noise is computed thanks to the correlation patterns in the detector plan discussed above.

\paragraph{NIRSpec IFU readout noise} -- 
Contrary to the NIRCam Imager detector, the plan of the NIRSpec IFU detector consists of a 1-D spatial and 1-D spectral image. More precisely, the optical system of NIRSpec IFU is composed of a slicer mirror array which slices the observed FOV into 0.1 arcsec-wide strips (corresponding to the NIRSpec IFU pixel size) to realign them in one dimension along one detector axis \citep{JDoxNIRSpecIFU}. For each spatial pixel, its spectrum is dispersed along the second detector axis. As a consequence, the corresponding readout noise is not independent from one spectral band to another. Thus, as previously explained above, this noise can be generated with a covariance matrix driven by the readout pattern features discussed above and added to the unfolded counterpart of the observed hyperspectral image after projection onto the detector plan.

\subsubsection{Zodiacal light, background and cosmic rays}

According to JWST documentation \citep{Kelsall1998,Pontoppidan2016}, the emissions from the Zodiacal cloud of the Solar System and of the Milky Way  as well as emission from the telescope are assumed to be negligible for bright sources, up to 5 microns. Furthermore, the JWST pipeline is designed to remove 99\% of cosmic rays impacts effects. These noise sources are thus neglected in this work. Note that a comprehensive model of background noise and cosmic rays impacts has been developed by the STScI for the ETC. 

\section{\label{expe}Experiments}

\subsection{Simulating observations using JWST forward models}

\begin{table*}
    \centering
        \caption{Properties of the simulated observed NIRCam Imager and NIRSpec IFU mosaic images, namely $\mathbf{Y}_\mathrm{m}$ and $\mathbf{Y}_\mathrm{h}$.}
    \begin{tabular}{lccc}
    \hline
    & \multicolumn{2}{c}{$\mathbf{Y}_\mathrm{m}$} & $\mathbf{Y}_\mathrm{h}$ \\
    \hline
    \hline
    Channel & SW & LW & \\
    \hline
    Wavelength range ($\mu$m)& 0.7 - 2.35 & 2.35 - 5.2 & 0.7 - 5.2\\
    Spectral points & 13 & 16 & 11586 \\
    \hline
    FOV (arcsec$^2$) & 30 $\times$ 30 & 30 $\times$ 30 & 30 $\times$ 30\\
    FOV (pixels$^2$) & 1000 $\times$ 1000 & 500 $\times$ 500 & 310 $\times$ 310 \\
    \hline
    \end{tabular}
    \label{tab:recap_simu_obs}
\end{table*}

\begin{figure*}
    \centering
    \begin{tikzpicture}
    
    \node[inner sep=0pt] (graph) at (0,0) {\includegraphics[width=0.35\linewidth]{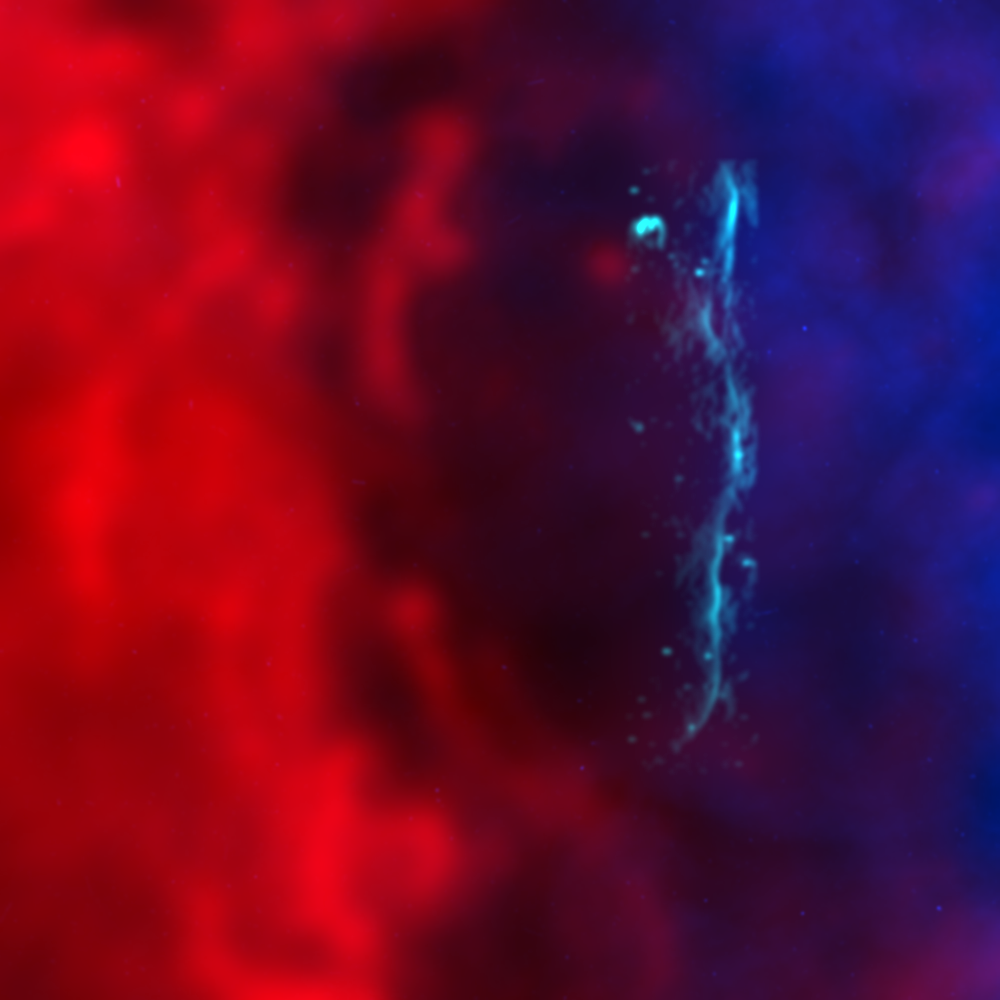}};
    \node[inner sep=0pt] (graph) at (0,-7) {\includegraphics[width=0.35\linewidth]{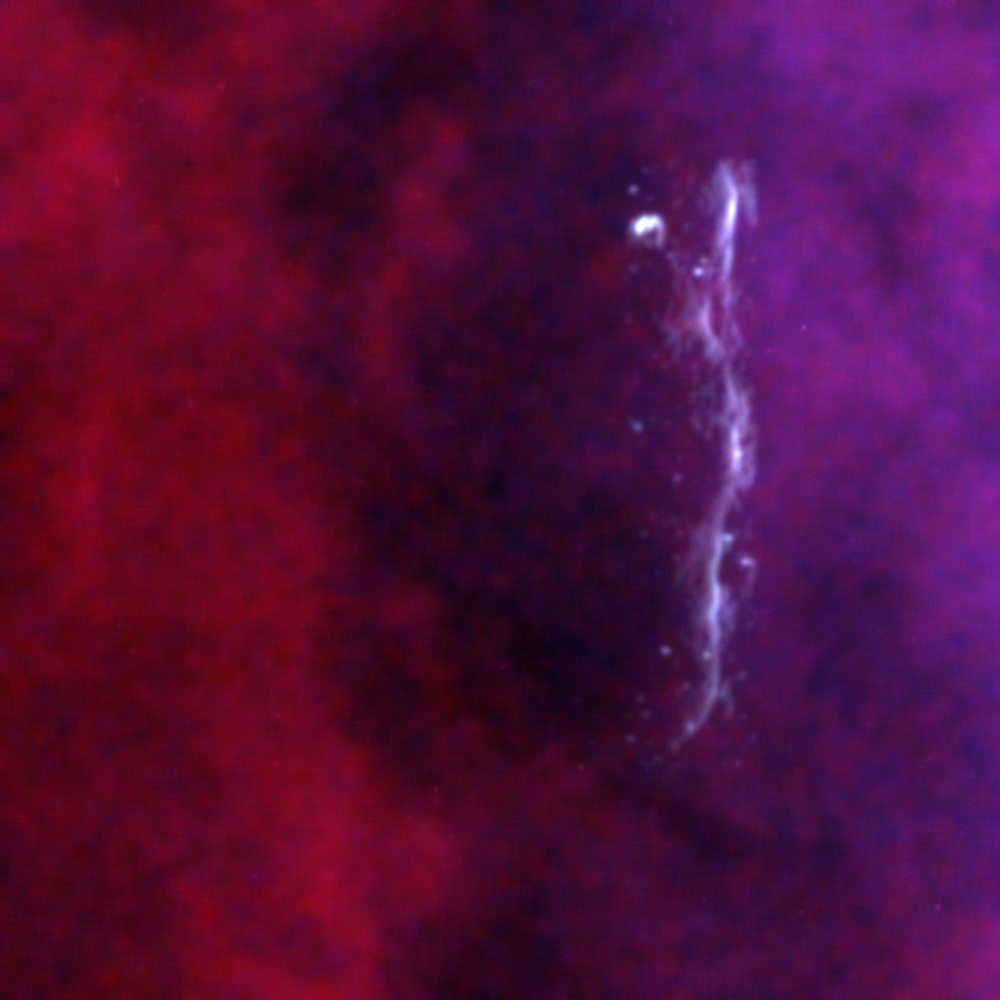}};
    \node[inner sep=0pt] (graph) at (0,-14) {\includegraphics[width=0.35\linewidth]{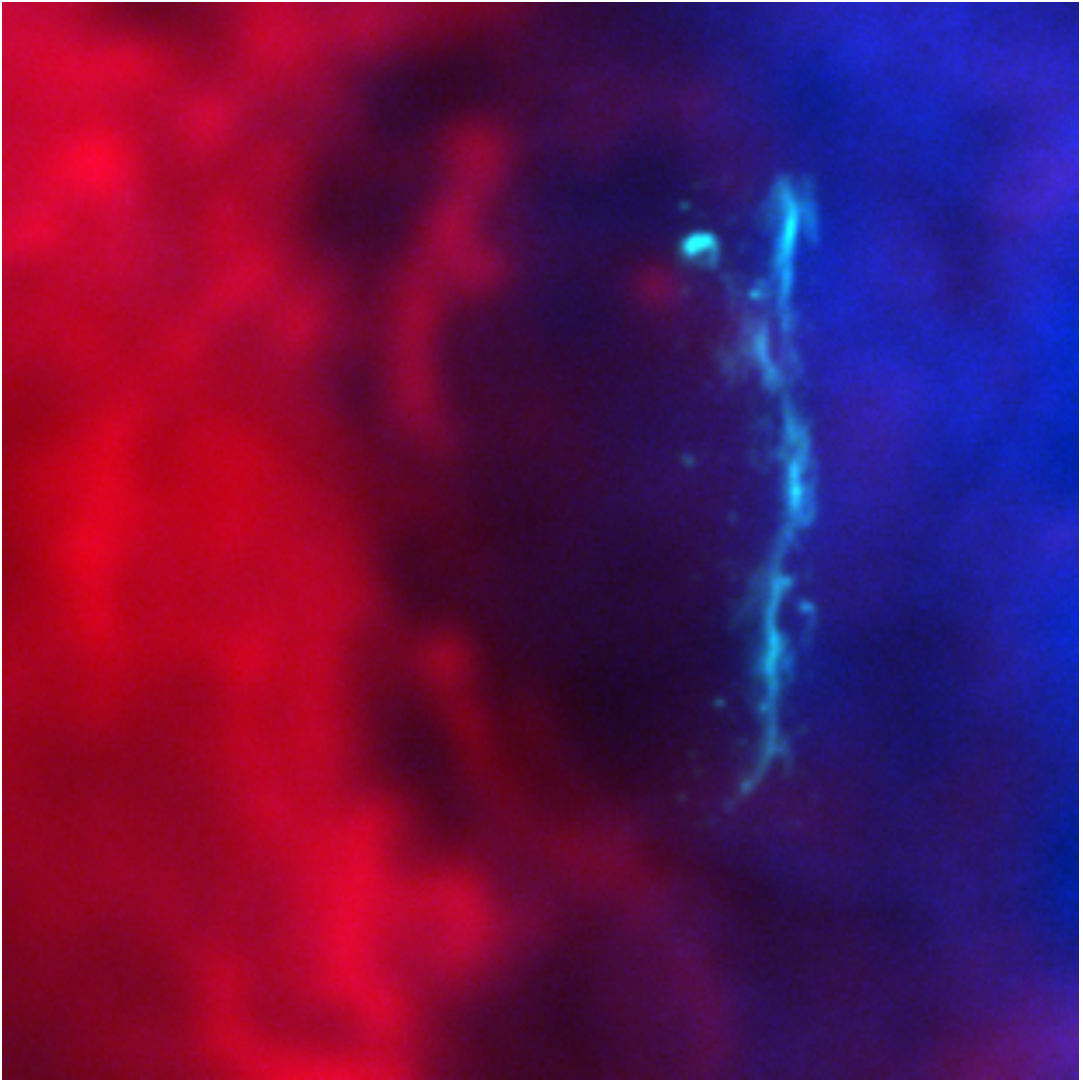}};
    
    \node[inner sep=0pt] (graph) at (9,0) {\includegraphics[width=0.54\linewidth]{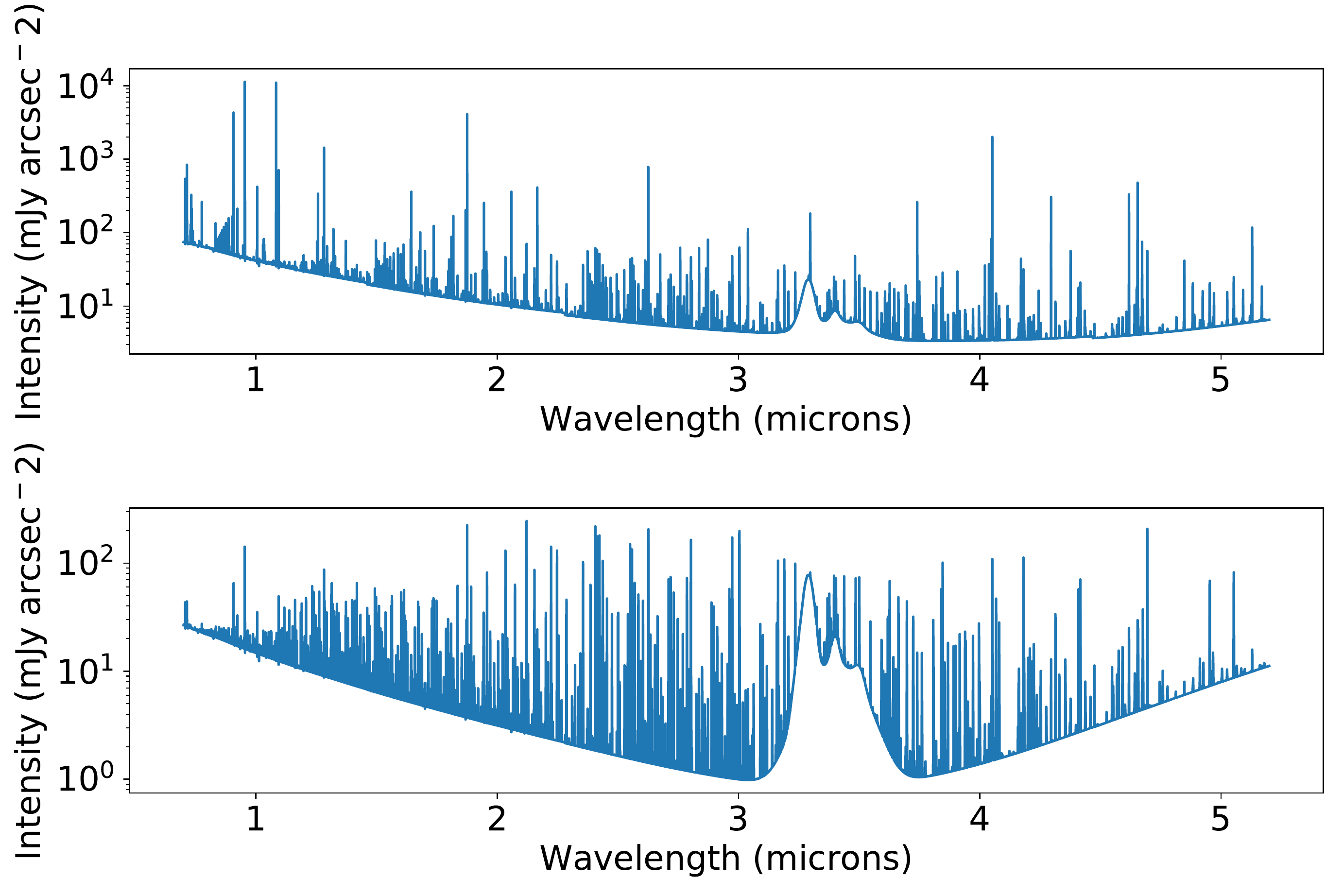}};
    \node[inner sep=0pt] (graph) at (9,-7) {\includegraphics[width=0.54\linewidth]{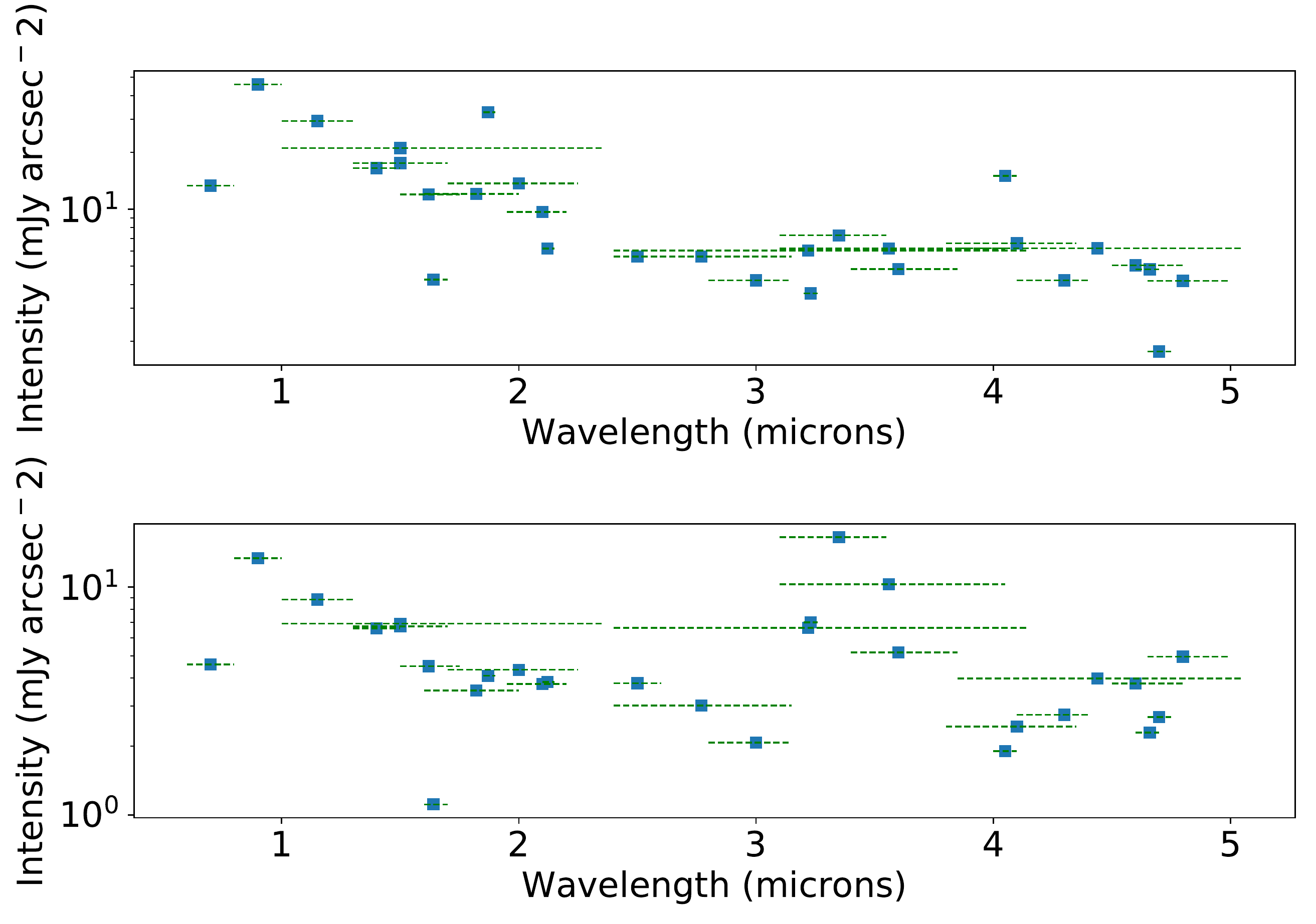}};
    \node[inner sep=0pt] (graph) at (9,-14) {\includegraphics[width=0.54\linewidth]{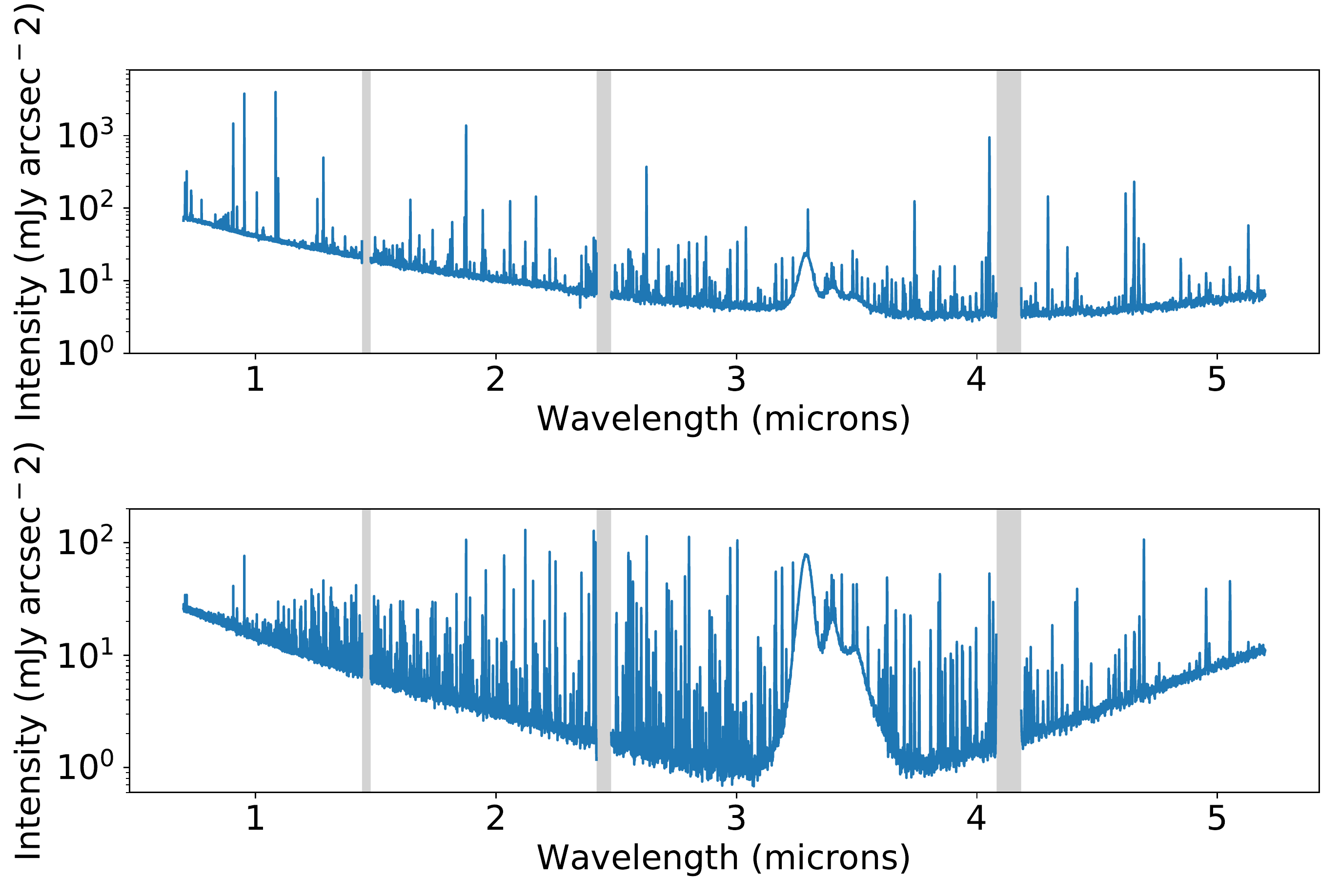}};
    
    \draw[orange, dashed] (1.45,2.1) -- ++(3.55,0);
    \node[orange] at (5,2.1) {$\bullet$};
    \node[orange] at (1.45,2.1) {$\bullet$};
    
    \draw[orange, dashed] (1.45,2.1-7) -- ++(3.6,0);
    \node[orange] at (5.05,2.1-7) {$\bullet$};
    \node[orange] at (1.45,2.1-7) {$\bullet$};
    
    \draw[orange, dashed] (1.45,2.1-14) -- ++(3.55,0);
    \node[orange] at (5,2.1-14) {$\bullet$};
    \node[orange] at (1.45,2.1-14) {$\bullet$};
    
    \draw[orange, dashed] (-1,-1.9) -- ++(6,0);
    \node[orange] at (-1,-1.9) {$\bullet$};
    \node[orange] at (5,-1.9) {$\bullet$};

    \draw[orange, dashed] (-1,-1.9-7) -- ++(6,0);
    \node[orange] at (-1,-1.9-7) {$\bullet$};
    \node[orange] at (5.05,-1.9-7) {$\bullet$};
    
    \draw[orange, dashed] (-1,-1.9-14) -- ++(6,0);
    \node[orange] at (-1,-1.9-14) {$\bullet$};
    \node[orange] at (5,-1.9-14) {$\bullet$};
    
    \node[thick] at (0,3.4) {\textbf{Synthetic scene}};
    \node[thick] at (0,-3.6) {\textbf{Simulated multispectral image}};
    \node[thick] at (0,-10.6) {\textbf{Simulated hyperspectral image}};
    
    \draw[orange,thick] (-3,1.6) rectangle ++(6,0.6);
    \draw[orange,thick] (-3,1.6-7) rectangle ++(6,0.6);
    \draw[orange,thick] (-3,1.6-14) rectangle ++(6,0.6);
    
    \draw[orange,thick] (5.4,0.75) rectangle ++(3.05,2);
    \draw[orange,thick] (5.4,0.75-3.25) rectangle ++(3.05,2);
    
    \draw[orange,thick] (5.4,0.75-6.95) rectangle ++(3.15,2.1);
    \draw[orange,thick] (5.4,0.75-6.95-3.42) rectangle ++(3.15,2.1);

    \draw[orange,thick] (5.4,0.75-14) rectangle ++(3.05,2);
    \draw[orange,thick] (5.4,0.75-17.25) rectangle ++(3.05,2);
    
    \end{tikzpicture}
    \caption{\textbf{Left} (from top to bottom)\textbf{:} RGB compositions of the synthetic PDR scene, the simulated NIRCam Imager multispectral image and the simulated NIRSpec IFU hyperspectral image. Red: H$_2$ emission line pic intensity at 2.122 $\mu$m (Narrow filter F212N for the multispectral image), Green: H recombination line pic intensity at 1.865 $\mu$m (Narrow filter F186N for the multispectral image), Blue: Fe$^+$ emission line pic intensity at 1.644 $\mu$m (Narrow filter F164N for the multispectral image). The observed field of view and spectral range considered in the fusion problem are represented by the orange boxes. \textbf{Right: } Two spectra from 0.7 to 5 microns associated to two pixels of each image on the left. From top to bottom, the first two are original spectra from the synthetic scene with about 12000 points, the following two are observed spectra from the multiband image provided by NIRCam Imager forward model with 29 spectral points, the last two are calibrated observed spectra from the hyperspectral image provided by NIRSpec IFU forward model with about 11000 points. Physical gaps in NIRSpec IFU detectors are specified in grey. Intensities are plotted in a logarithmic scale. The $1$st, $3$rd and $5$th spectra are dominated by ionization front and ionized region emissions while the $2$nd, $4$th and $6$th are dominated by dissociation front and molecular cloud emissions.}
    \label{fig:simus}
\end{figure*}
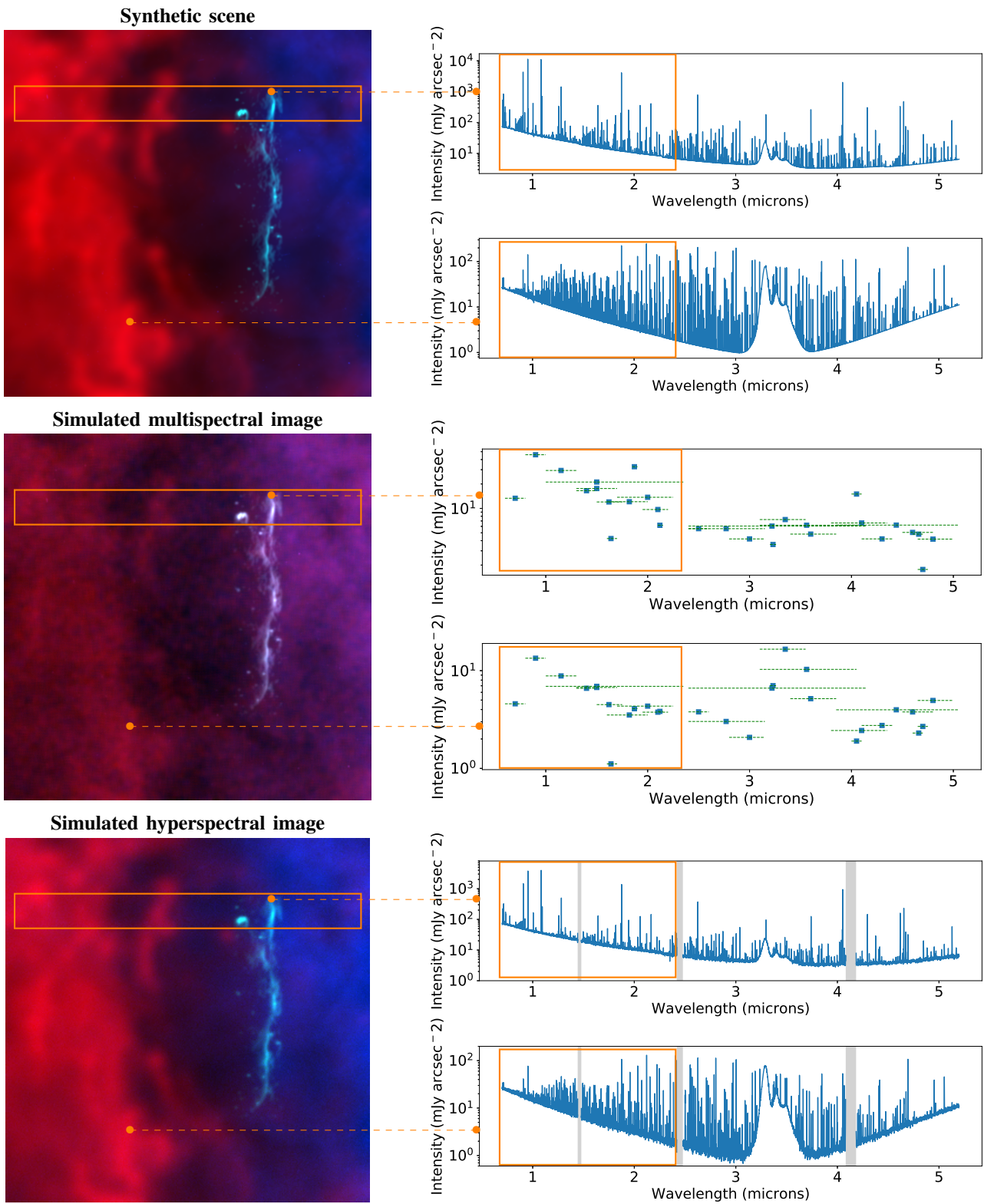

This section capitalizes on the forward models of NIRCam Imager and NIRSpec IFU and the associated noise model proposed in Sect. \ref{sec:forward} to simulate observations associated with the synthetic astrophysical scene generated according to the framework introduced in Sect.~\ref{synthpdr}. 
To adjust the characteristics of the noise, we rely on the integration times as planned by the ERS 1288 program of \cite{Berne2017}. The observing parameters of this program can be downloaded publicly through the Astronomer's Proposal Tool (APT) provided by the STScI.  
The characteristics of the resulting simulated multispectral and hyperspectral images, respectively denoted as $\mathbf{Y}_\mathrm{m}$ and $\mathbf{Y}_\mathrm{h}$, are summarized in Tab.~\ref{tab:recap_simu_obs}.
The FOV of the resulting simulated image $\mathbf{Y}_{\mathrm{m}}$ corresponds to about one sixteenth of the total NIRCam Imager FOV since the  synthetic scene is smaller than the actual full NIRCam Imager FOV. On the other hand, the FOV of the simulated hyperspectral image $\mathbf{Y}_{\mathrm{h}}$ corresponds to a mosaic of 10 $\times$ 10 NIRSpec IFU FOVs.
These simulated multispectral and hyperspectral images are shown in Fig. \ref{fig:simus}. To illustrate the contents of the simulated dataset, we present red-green-blue (RGB) colored compositions of the images as well as spectra extracted at specific positions, for the scene and simulated observations (see Fig.~\ref{fig:simus} for details of the composition). The simulations show how the instruments degrade the spectral and spatial resolution of the fully resolved synthetic astrophysical scene. More precisely, for the multispectral observations, the RGB composition shows less contrast, due to the loss of spectral information due to the wide filters. The hyperspectral data is clearly less spatially resolved, and the spectra exhibit a high level of noise. Overall, for the considered realistic scene of the Orion Bar which is a bright source, it is worth noting that the signal-to-noise ratio remains high for most parts of the images and spectra.


\subsection{Fusion of simulated observations}
\subsubsection{Method}
\label{subsec:fusion}
The synthetic scene and the simulated observed NIRCam Imager and NIRSpec IFU images have been generated to assess the performance of a dedicated fusion method we have developed \citep{Guilloteau2019}. We refer the reader to this latter paper for full details about the method, but provide below the main characteristics of the fusion algorithm. 
The fusion task is formulated as an inverse problem, relying on the forward models specifically developed for the JWST NIRCam Imager and NIRSpec IFU instruments in Sec. \ref{sec:forward}. More precisely, the fused product $\hat{\mathbf{X}}$ is defined as a minimizer of the objective function $\mathcal{J}(\cdot)$ given by
\begin{multline}
    \mathcal{J}(\mathbf{X}) = \frac{1}{2\sigma_\mathrm{m}^2} \left\| \mathbf{Y}_\mathrm{m} - \mathbf{L}_\mathrm{m} \mathcal{M}(\mathbf{X}) \right\|_{\mathrm{F}}^2 + \frac{1}{2\sigma_\mathrm{h}^2} \left\| \mathbf{Y}_\mathrm{h} - \mathbf{L}_\mathrm{h} \mathcal{H}(\mathbf{X}) \mathbf{S} \right\|_{\mathrm{F}}^2 \\ 
    + \varphi_\mathrm{spe} (\mathbf{X}) + \varphi_\mathrm{spa} (\mathbf{X})
    \label{pinv}
\end{multline}
where $\|\cdot\|_{\mathrm{F}}$ is the Frobenius norm.
The two first terms are referred to as data fidelity terms and $\sigma_\mathrm{m}^2$ and $\sigma_\mathrm{h}^2$ are their respective weights associated to the the noise level in each observed image $\mathbf{Y}_\mathrm{m}$ and $\mathbf{Y}_\mathrm{h}$. The noisier the greater $\sigma_\mathrm{m}^2$ or $\sigma_\mathrm{h}^2$, and the less significant the related data fidelity term. The complementary terms $\varphi_\mathrm{spe}(\mathbf{X})$ and $\varphi_\mathrm{spa}(\mathbf{X})$ are respectively spectral and spatial regularizations summarizing {\it{a priori}} information on the expected fused image. 
In the approach advocated by \cite{Guilloteau2019}, the spectral regularization $\varphi_\mathrm{spa}(\cdot)$ in \eqref{pinv}  relies on the prior assumption that the spectra of the fused image live in a low dimensional subspace whereas the spatial regularization $\varphi_\mathrm{spa}(\cdot)$ promotes a smooth spatial content. Due to the high-dimensionality of the resulting optimization problem, its solution cannot be analytically computed but requires an iterative procedure. To get a scalable and fast algorithm able to handle realistic measurements, \cite {Guilloteau2019} proposed two computational tricks: {\it i)} the problem is formulated in the Fourier domain, where the convolution operators $\mathcal{H}(\cdot)$ and $\mathcal{M}(\cdot)$ can be efficiently implemented and {\it ii)} in a preprocessing step, the JWST forward models are computed in the lower-dimensional subspace induced by the spectral regularization, which leads to sparse and easily storable operators. By combining these two tricks, the final algorithmic procedure minimizing $\mathcal{J}(\cdot)$ saves about 90\% of the computational time with respect to a naive implementation.

\subsubsection{Results}

In this work, we perform the fusion task on a subset of the simulated multi- and hyperspectral observed images. This choice has been first guided by the observing strategy currently considered in the ERS 1288 observing program  for the Orion Bar \citep{Berne2017}. In practice, as depicted in Fig.~\ref{fig:simus}, the FOV used for fusion (orange boxes in the right-hand side) is limited to a 2.7 $\times$ 27 arcsec$^2$ cut across the Bar, representing a mosaic of 9 NIRSpec IFUs FOVs. Secondly, as the SW and LW channels of the NIRCam Imager present distinct spatial sampling properties, we restrict the test of the fusion algorithm to the spectral range of the shorter wavelengths, between 0.7 and 2.35 $\mu$m, where the ratio of spatial resolution between imager and spectrometer is largest (i.e. where the fusion is most difficult).  In the end, the objective is to fuse a 13 $\times$ 90 $\times$ 900 pixels simulated multispectral image and a 5000 $\times$ 28 $\times$ 279 pixels simulated hyperspectral image.

\begin{figure*}
    \centering
    \begin{tikzpicture}
    \node[inner sep=0pt] (graph) at (0,0) {\includegraphics[width=0.3\linewidth]{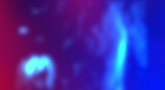}};
    \node[inner sep=0pt] (graph) at (6,0) {\includegraphics[width=0.29\linewidth]{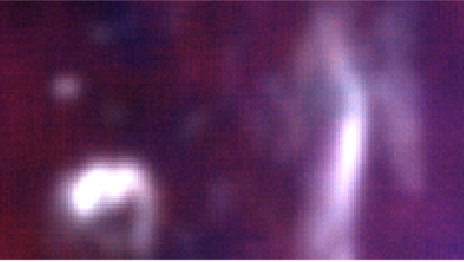}};
    \node[inner sep=0pt] (graph) at (0,-4) {\includegraphics[width=0.3\linewidth]{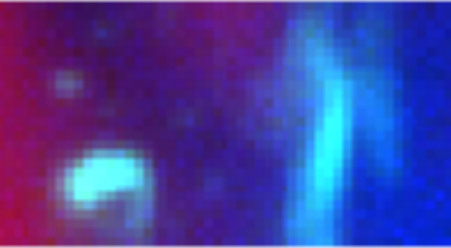}};
    \node[inner sep=0pt] (graph) at (6,-4) {\includegraphics[width=0.3\linewidth]{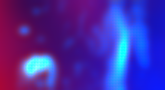}};
    
    \node[thick] at (0,1.8) {\textbf{Synthetic scene}};
    \node[thick] at (6,1.8) {\textbf{Simulated multispectral image}};
    \node[thick] at (0,1.8-4) {\textbf{Simulated hyperspectral image}};
    \node[thick] at (6,1.8-4) {\textbf{Reconstructed image}};
    \end{tikzpicture}
    \caption{Zoom on ionization front strong structures from Fig. \ref{fus_res}.}
    \label{zoom}
\end{figure*}

The fused image has been obtained after about 2000 seconds of pre-processing (dedicated to the pre-computation of the JWST forward models in the lower dimensional subspace) and 20 seconds of iterative minimization of the objective function $\mathcal{J}(\cdot)$. Qualitatively and generally speaking, the reconstruction is excellent from spectral and spatial points-of-view. Regarding the spectra, the fusion is very good for pixels which are located on smooth spatial structures. Efficient denoising can be observed since reconstructed spectra show much less noise than the simulated NIRSpec IFU hyperspectral image. However, in regions with significant and sharp variations of the intensity at small spatial scales (such as the ionization front), the fusion procedure appears to be less accurate. This is likely due to the chosen spatial regularization which tends to promote smooth images, and therefore distributes the flux over neighboring pixels. This issue is currently under investigation to provide a better regularization able to mitigate this effect. Similar conclusions can be drawn when analyzing the spatial content of the fused image, as  illustrated in Fig.~\ref{zoom}. Overall, the reconstruction is very good, and a significant denoising is also observed. The gain in resolution of the reconstructed image with respect to the hyperspectral image is clearly noticeable, but thin structures, such as the ionization front, are smoother than in the original simulated astrophysical scene. Again, this is likely due to the regularization. 

We now turn to a more quantitative analysis of the performance of the fusion method. To do so, we consider the reconstruction SNR of the fused image $\overline{\mathbf{X}}$ with respect to the corresponding actual scene $\mathbf{X}$. It is expressed as 
\begin{equation}
\mathrm{SNR} = 10\log_{10} \left( \frac{\left\|\mathbf{X}\right\|_2^2}{\left\| \mathbf{X}-\overline{\mathbf{X}} \right\|_{\mathrm{F}}^2}\right) 
\end{equation}
The reconstruction SNR reached by the proposed fusion procedure is compared to the SNR associated with an up-sampled counterpart of the observed hyperspectral image obtained by a simple band-wise bi-cubic spline interpolation to the resolution of the synthetic scene. The resulting SNRs are respectively 18.5 and 10.6 for the fused product and the up-sampled observed hyperspectral image. This means that spatial and spectral contents are much more accurately reconstructed by the fusion process proposed by \cite{Guilloteau2019}. Such results underline the benefit of data fusion compared with considering only the observed hyperspectral image, discarding the information brought by the multispectral image.

\subsubsection{Perspectives for fusion methods in the context of the JWST mission}
Overall, the results of data fusion performed on simulated multispectral and hyperspectral JWST  images show high quality spectral and spatial reconstruction of the scene. Most of spectral and spatial details lost either in the multispectral or in the hyperspectral image are recovered in the fused product. Such results supports further investigations on this fusion method, with great promises of application on real data. One preliminary step to be achieved concerns a more comprehensive performance assessment. It is indeed necessary to evaluate the benefit of the the fusion procedure when dealing with simulated observations obtained from the from JWST scientific team through the Exposure Time Calculator (ETC) tool, with possibly the synthesis of the same 3D complex scene described in Sec. \ref{synthpdr}. This is a project that we are currently undertaking with STScI. Current limitations of the methods we have identified concern the unsatisfactory reconstruction of sharp structures in the synthetic scene, due to the chosen spatial regularization, which promotes a smooth spatial content in the fused product. Considering a regularization term in the objective function $\mathcal{J}(\cdot)$ defined in \eqref{pinv} is necessary not to over-fit the noise in the observed images. Future works should address this issue by designing a tailored regularization.

\section{\label{ccl}Conclusion}

In this work we built a synthetic scene of a photodissociation region located in the Orion Bar with high spatio-spectral resolution. This scene has been created according to current models with simulated spectra and spatial maps derived from real data. Forward models of two instruments embedded on the JWST, namely NIRCam Imager and NIRSpec IFU, were developed and used to simulate JWST observations of the Orion Bar PDR. These simulated data were used to assess the performance of a fusion method we developed. The results showed to be promising  allows for recovering  spectroscopic and spatial details which were lost in the simulated NIRCam Imager and NIRSpec IFU  observations.  This suggested that image fusion of JWST data would offer a significant enhancement of scientific interpretation. However, improvements of the fusion method are still required, in particular to mitigate effect of the regularization. Tests on synthetic data with a more realistic noise than the one considered in this paper are also necessary. 

\section*{Acknowledgements}
The authors thank K. Pontoppidan for his feedback on the ETC and the JWST documentation, and the members of the core team of ERS project ``Radiative Feedback from Massive Stars as Traced by Multiband Imaging and Spectroscopic Mosaics'' for providing the spectra used to create the simulated datasets. They are also thankful to A. Albergel, F. Orieux and R. Abi Rizk for fruitful discussions regarding this work.
Part of this work has been supported by the ANR-3IA Artificial and 
Natural Intelligence Toulouse Institute (ANITI), the French
Programme Physique et Chimie du Milieu Interstellaire (PCMI) funded by the
Conseil  National  de  la  Recherche  Scientifique  (CNRS)  and  Centre  National
d'\'Etudes Spatiales (CNES).

\bibliographystyle{IEEEtranN}
\bibliography{strings_all_ref,refs}

\begin{thebibliography}{43}
\providecommand{\natexlab}[1]{#1}
\providecommand{\url}[1]{#1}
\csname url@samestyle\endcsname
\providecommand{\newblock}{\relax}
\providecommand{\bibinfo}[2]{#2}
\providecommand{\BIBentrySTDinterwordspacing}{\spaceskip=0pt\relax}
\providecommand{\BIBentryALTinterwordstretchfactor}{4}
\providecommand{\BIBentryALTinterwordspacing}{\spaceskip=\fontdimen2\font plus
\BIBentryALTinterwordstretchfactor\fontdimen3\font minus
  \fontdimen4\font\relax}
\providecommand{\BIBforeignlanguage}[2]{{%
\expandafter\ifx\csname l@#1\endcsname\relax
\typeout{** WARNING: IEEEtranN.bst: No hyphenation pattern has been}%
\typeout{** loaded for the language `#1'. Using the pattern for}%
\typeout{** the default language instead.}%
\else
\language=\csname l@#1\endcsname
\fi
#2}}
\providecommand{\BIBdecl}{\relax}
\BIBdecl

\bibitem[{Gardner} et~al.(2006){Gardner}, {Mather}, {Clampin}, {Doyon},
  {Greenhouse}, {Hammel}, {Hutchings}, {Jakobsen}, {Lilly}, and
  {Long}]{Gardner2006}
J.~P. {Gardner} \emph{et~al.}, ``{The James Webb Space Telescope},''
  \emph{Space Science Reviews}, vol. 123, no.~4, pp. 485--606, Apr 2006.

\bibitem[{Rieke} et~al.(2005){Rieke}, {Kelly}, and {Horner}]{Rieke2005}
M.~J. {Rieke}, D.~{Kelly}, and S.~{Horner}, ``{Overview of James Webb Space
  Telescope and NIRCam's Role},'' in \emph{Cryogenic Optical Systems and
  Instruments XI}, ser. Society of Photo-Optical Instrumentation Engineers
  (SPIE) Conference Series, J.~B. {Heaney} and L.~G. {Burriesci}, Eds., vol.
  5904, Aug 2005, pp. 1--8.

\bibitem[{Bagnasco} et~al.(2007){Bagnasco}, {Kolm}, {Ferruit}, {Honnen},
  {Koehler}, {Lemke}, {Maschmann}, {Melf}, {Noyer}, and {Rumler}]{Bagnasco2007}
G.~{Bagnasco} \emph{et~al.}, ``{Overview of the near-infrared spectrograph
  (NIRSpec) instrument on-board the James Webb Space Telescope (JWST)},'' in
  \emph{Cryogenic Optical Systems and Instruments XII}, ser. Society of
  Photo-Optical Instrumentation Engineers (SPIE) Conference Series, vol. 6692,
  Sep 2007, p. 66920M.

\bibitem[{Doyon} et~al.(2012){Doyon}, {Hutchings}, {Beaulieu}, {Albert},
  {Lafreni{\`e}re}, {Willott}, {Touahri}, {Rowlands}, {Maszkiewicz}, and
  {Fullerton}]{Doyon2012}
R.~{Doyon} \emph{et~al.}, ``{The JWST Fine Guidance Sensor (FGS) and
  Near-Infrared Imager and Slitless Spectrograph (NIRISS)},'' in \emph{Space
  Telescopes and Instrumentation 2012: Optical, Infrared, and Millimeter Wave},
  ser. Society of Photo-Optical Instrumentation Engineers (SPIE) Conference
  Series, vol. 8442, Sep 2012, p. 84422R.

\bibitem[{Rieke} et~al.(2015){Rieke}, {Wright}, {B{\"o}ker}, {Bouwman},
  {Colina}, {Glasse}, {Gordon}, {Greene}, {G{\"u}del}, and
  {Henning}]{Rieke2015}
G.~H. {Rieke} \emph{et~al.}, ``The mid-infrared instrument for the {James Webb
  Space Telescope, I: Introduction},'' \emph{Publ. Astron. Society of the
  Pacific}, vol. 127, no. 953, p. 584, Jul 2015.

\bibitem[{Wei} et~al.(2015){Wei}, {Dobigeon}, and {Tourneret}]{Wei2015}
Q.~{Wei}, N.~{Dobigeon}, and J.-Y. {Tourneret}, ``Fast fusion of multi-band
  images based on solving a {S}ylvester equation,'' \emph{IEEE Trans. Image
  Process.}, vol.~24, no.~11, pp. 4109--4121, Nov 2015.

\bibitem[{Simoes} et~al.(2015){Simoes}, {Bioucas-Dias}, {Almeida}, and
  {Chanussot}]{Simoes2015}
M.~{Simoes}, J.~{Bioucas-Dias}, L.~B. {Almeida}, and J.~{Chanussot}, ``A convex
  formulation for hyperspectral image superresolution via subspace-based
  regularization,'' \emph{IEEE Trans. Geosci. Remote Sens.}, vol.~53, no.~6,
  pp. 3373--3388, Jun 2015.

\bibitem[{Yokoya} et~al.(2012){Yokoya}, {Yairi}, and {Iwasaki}]{Yokoya2012}
N.~{Yokoya}, T.~{Yairi}, and A.~{Iwasaki}, ``Coupled nonnegative matrix
  factorization unmixing for hyperspectral and multispectral data fusion,''
  \emph{IEEE Trans. Geosci. Remote Sens.}, vol.~50, no.~2, pp. 528--537, Feb
  2012.

\bibitem[Wald et~al.(2005)Wald, Ranchin, and Mangolini]{Wald2005}
L.~Wald, T.~Ranchin, and M.~Mangolini, ``Fusion of satellite images of
  different spatial resolutions: assessing the quality of resulting image,''
  \emph{IEEE Trans. Geosci. Remote Sensing}, vol.~43, pp. 1391--1402, 2005.

\bibitem[{Bern\'e} et~al.(2017){Bern\'e}, {Habart}, {Peeters}, {Abergel},
  {Bergin}, {Bernard-Salas}, {Bron}, {Cami}, {Cazaux}, and
  {Dartois}]{Berne2017}
O.~{Bern\'e} \emph{et~al.}, ``Radiative feedback from massive stars as traced
  by multiband imaging and spectroscopic mosaics,'' JWST Proposal ID 1288.
  Cycle 0 Early Release Science, p. 1288, Nov 2017.

\bibitem[Tielens and Hollenbach(1985)]{Tielens1985}
A.~Tielens and D.~Hollenbach, ``Photodissociation regions. {I-Basic} model.
  {II-A} model for the orion photodissociation region,'' \emph{Astrophys. J.},
  vol. 291, pp. 722--754, 1985.

\bibitem[Adams et~al.(2004)Adams, Hollenbach, Laughlin, and Gorti]{Adams2004}
F.~C. Adams, D.~Hollenbach, G.~Laughlin, and U.~Gorti, ``Photoevaporation of
  circumstellar disks due to external far-ultraviolet radiation in stellar
  aggregates,'' \emph{The Astrophysical Journal}, vol. 611, no.~1, p. 360,
  2004.

\bibitem[Gorti and Hollenbach(2008)]{Gorti2008}
U.~Gorti and D.~Hollenbach, ``Photoevaporation of circumstellar disks by
  far-ultraviolet, extreme-ultraviolet and x-ray radiation from the central
  star,'' \emph{The Astrophysical Journal}, vol. 690, no.~2, p. 1539, 2008.

\bibitem[{Champion} et~al.(2017){Champion}, {Bern{\'e}}, {Vicente}, {Kamp}, {Le
  Petit}, {Gusdorf}, {Joblin}, and {Goicoechea}]{Champion2017}
J.~{Champion} \emph{et~al.}, ``{Herschel survey and modelling of
  externally-illuminated photoevaporating protoplanetary disks},''
  \emph{Astron. \& Astrophys.}, vol. 604, p. A69, Aug 2017.

\bibitem[{Bernard-Salas} and {Tielens}(2005)]{Bernard2005}
J.~{Bernard-Salas} and A.~G.~G.~M. {Tielens}, ``Physical conditions in
  photo-dissociation regions around planetary nebulae,'' \emph{Astron. \&
  Astrophys.}, vol. 431, pp. 523--538, Feb. 2005.

\bibitem[{Cox} et~al.(2016){Cox}, {Pilleri}, {Bern{\'e}}, {Cernicharo}, and
  {Joblin}]{Cox2016}
N.~L.~J. {Cox} \emph{et~al.}, ``Polycyclic aromatic hydrocarbons and molecular
  hydrogen in oxygen-rich planetary nebulae: the case of {NGC} 6720,''
  \emph{Monthly Notices of the Royal Astron. Society : Letters}, vol. 456,
  no.~1, pp. L89--L93, Feb 2016.

\bibitem[{Tielens}(2005)]{Tielens2005}
A.~G.~G.~M. {Tielens}, \emph{The Physics and Chemistry of the Interstellar
  Medium}.\hskip 1em plus 0.5em minus 0.4em\relax Cambridge Univ. Press, 2005.

\bibitem[{Goicoechea} et~al.(2016){Goicoechea}, {Pety}, {Cuadrado},
  {Cernicharo}, {Chapillon}, {Fuente}, {Gerin}, {Joblin}, {Marcelino}, and
  {Pilleri}]{Goicoechea2016}
J.~R. {Goicoechea} \emph{et~al.}, ``{Compression and ablation of the
  photo-irradiated molecular cloud the Orion Bar},'' \emph{Nature}, vol. 537,
  no. 7619, pp. 207--209, Sep 2016.

\bibitem[{Joblin} et~al.(2018){Joblin}, {Bron}, {Pinto}, {Pilleri}, {Le Petit},
  {Gerin}, {Le Bourlot}, {Fuente}, {Berné}, and {Goicoechea}]{Joblin2018}
C.~{Joblin} \emph{et~al.}, ``{Structure of photodissociation fronts in
  star-forming regions revealed by Herschel observations of high-J CO emission
  lines},'' \emph{Astron. \& Astrophys.}, vol. 615, p. A129, Jul 2018.

\bibitem[{Fuente} et~al.(2005){Fuente}, {Garc{\'\i}a-Burillo}, {Gerin},
  {Teyssier}, {Usero}, {Rizzo}, and {de Vicente}]{Fuente2005}
A.~{Fuente} \emph{et~al.}, ``Photon-dominated chemistry in the nucleus of
  {M}82: {W}idespread {HOC$^{+}$} emission in the inner 650 parsec disk,''
  \emph{The Astrophysical J. Lett.}, vol. 619, no.~2, pp. L155--L158, Feb 2005.

\bibitem[Habart et~al.(2011)Habart, Abergel, Boulanger, Joblin, Verstraete,
  Compi{\`e}gne, des For{\^e}ts, and Le~Bourlot]{Habart2011}
E.~Habart \emph{et~al.}, ``Excitation of h2 in photodissociation regions as
  seen by spitzer,'' \emph{Astron. \& Astrophys.}, vol. 527, p. A122, 2011.

\bibitem[{Bron} et~al.(2014){Bron}, {Le Bourlot}, and {Le Petit}]{Bron2014}
E.~{Bron}, J.~{Le Bourlot}, and F.~{Le Petit}, ``{Surface chemistry in the
  interstellar medium. II. H$_{2}$ formation on dust with random temperature
  fluctuations},'' \emph{Astron. \& Astrophys.}, vol. 569, p. A100, Sep 2014.

\bibitem[R{\"o}llig et~al.(2007)R{\"o}llig, Abel, Bell, Bensch, Black, Ferland,
  Jonkheid, Kamp, Kaufman, Le~Bourlot, et~al.]{Roellig2007}
M.~R{\"o}llig \emph{et~al.}, ``A photon dominated region code comparison
  study,'' \emph{Astron. \& Astrophys.}, vol. 467, no.~1, pp. 187--206, 2007.

\bibitem[{Bern{\'e}} et~al.(2015){Bern{\'e}}, {Montillaud}, and
  {Joblin}]{Berne2015}
O.~{Bern{\'e}}, J.~{Montillaud}, and C.~{Joblin}, ``Top-down formation of
  fullerenes in the interstellar medium,'' \emph{Astron. \& Astrophys.}, vol.
  577, p. A133, May 2015.

\bibitem[{Peeters} et~al.(2017){Peeters}, {Bauschlicher}, {Allamand ola},
  {Tielens}, {Ricca}, and {Wolfire}]{Peeters2017}
E.~{Peeters} \emph{et~al.}, ``The {PAH} emission characteristics of the
  reflection {N}ebula {NGC} 2023,'' \emph{Astrophys. J.}, vol. 836, no.~2, p.
  198, Feb 2017.

\bibitem[{Bally}(2015)]{Bally2015}
J.~e.~a. {Bally}, ``The first ultraviolet survey of {Orion Nebula's}
  protoplanetary disks and outflows,'' HST Proposal ID 13419. Cycle 21, 2015.

\bibitem[{Fazio} et~al.(2004){Fazio}, {Hora}, {Allen}, {Ashby}, {Barmby},
  {Deutsch}, {Huang}, {Kleiner}, {Marengo}, {Megeath}, {Melnick}, {Pahre},
  {Patten}, {Polizotti}, {Smith}, {Taylor}, {Wang}, {Willner}, {Hoffmann},
  {Pipher}, {Forrest}, {McMurty}, {McCreight}, {McKelvey}, {McMurray}, {Koch},
  {Moseley}, {Arendt}, {Mentzell}, {Marx}, {Losch}, {Mayman}, {Eichhorn},
  {Krebs}, {Jhabvala}, {Gezari}, {Fixsen}, {Flores}, {Shakoorzadeh}, {Jungo},
  {Hakun}, {Workman}, {Karpati}, {Kichak}, {Whitley}, {Mann}, {Tollestrup},
  {Eisenhardt}, {Stern}, {Gorjian}, {Bhattacharya}, {Carey}, {Nelson},
  {Glaccum}, {Lacy}, {Lowrance}, {Laine}, {Reach}, {Stauffer}, {Surace},
  {Wilson}, {Wright}, {Hoffman}, {Domingo}, and {Cohen}]{Fazio2004}
G.~G. {Fazio} \emph{et~al.}, ``{The Infrared Array Camera (IRAC) for the
  Spitzer Space Telescope},'' \emph{The Astrophysical J. Supplement Series},
  vol. 154, no.~1, pp. 10--17, Sep 2004.

\bibitem[{Weingartner} and {Draine}(2001)]{Weingartner2001}
J.~C. {Weingartner} and B.~T. {Draine}, ``Dust grain-size distributions and
  extinction in the {Milky Way}, large magellanic cloud, and small magellanic
  cloud,'' \emph{Astrophys. J.}, vol. 548, no.~1, pp. 296--309, Feb 2001.

\bibitem[{Le Petit} et~al.(2006){Le Petit}, {Nehm{\'e}}, {Le Bourlot}, and
  {Roueff}]{Lepetit2006}
F.~{Le Petit}, C.~{Nehm{\'e}}, J.~{Le Bourlot}, and E.~{Roueff}, ``A model for
  atomic and molecular interstellar gas: {T}he {Meudon PDR} code,'' \emph{The
  Astrophysical J. Supplement Series}, vol. 164, pp. 506--529, June 2006.

\bibitem[{Ferland} et~al.(1998){Ferland}, {Korista}, {Verner}, {Ferguson},
  {Kingdon}, and {Verner}]{Ferland1998}
G.~J. {Ferland} \emph{et~al.}, ``{CLOUDY} 90: Numerical simulation of plasmas
  and their spectra,'' \emph{Publ. Astron. Society of the Pacific}, vol. 110,
  pp. 761--778, July 1998.

\bibitem[{Pilleri} et~al.(2012){Pilleri}, {Montillaud}, {Bern{\'e}}, and
  {Joblin}]{Pilleri2012}
P.~{Pilleri}, J.~{Montillaud}, O.~{Bern{\'e}}, and C.~{Joblin}, ``{Evaporating
  very small grains as tracers of the UV radiation field in photo-dissociation
  regions},'' \emph{Astron. \& Astrophys.}, vol. 542, p. A69, June 2012.

\bibitem[{Compi{\`e}gne} et~al.(2011){Compi{\`e}gne}, {Verstraete}, {Jones},
  {Bernard}, {Boulanger}, {Flagey}, {Le Bourlot}, {Paradis}, and
  {Ysard}]{Compiegne2011}
M.~{Compi{\`e}gne} \emph{et~al.}, ``The global dust {SED}: tracing the nature
  and evolution of dust with {DustEM},'' \emph{Astron. \& Astrophys.}, vol.
  525, p. A103, Jan. 2011.

\bibitem[{Damelin} and {Hoang}(2018)]{Damelin2017}
S.~B. {Damelin} and N.~S. {Hoang}, ``On surface completion and image inpainting
  by biharmonic functions: Numerical aspects,'' \emph{Int. J. Mathematics and
  Mathematical Sciences}, Feb. 2018.

\bibitem[{Pontoppidan} et~al.(2016){Pontoppidan}, {Pickering}, {Laidler},
  {Gilbert}, {Sontag}, {Slocum}, {Sienkiewicz}, {Hanley}, {Earl}, and
  {Pueyo}]{Pontoppidan2016}
K.~M. {Pontoppidan} \emph{et~al.}, ``{Pandeia: a multi-mission exposure time
  calculator for JWST and WFIRST},'' in \emph{Observatory Operations:
  Strategies, Processes, and Systems VI}, ser. Society of Photo-Optical
  Instrumentation Engineers (SPIE) Conference Series, vol. 9910, Jul 2016, p.
  991016.

\bibitem[{Perrin} et~al.(2012){Perrin}, {Soummer}, {Elliott}, {Lallo}, and
  {Sivaramakrishnan}]{Perrin2012}
M.~D. {Perrin} \emph{et~al.}, ``{Simulating point spread functions for the
  James Webb Space Telescope with WebbPSF},'' in \emph{Space Telescopes and
  Instrumentation 2012: Optical, Infrared, and Millimeter Wave}, ser. Society
  of Photo-Optical Instrumentation Engineers (SPIE) Conference Series, vol.
  8442, Sep 2012, p. 84423D.

\bibitem[{Space Telescope Science Institute
  (STScI)}(2017{\natexlab{a}})]{JDoxNIRCamFilters}
\BIBentryALTinterwordspacing
{Space Telescope Science Institute (STScI)}, \emph{NIRCam Filters, JWST User
  Documentation [Published 2017 November 29]}, Baltimore, MD, 2017. [Online].
  Available:
  \url{https://jwst-docs.stsci.edu/near-infrared-camera/nircam-instrumentation/nircam-filters}
\BIBentrySTDinterwordspacing

\bibitem[Hadj-Youcef et~al.(2018)Hadj-Youcef, Orieux, Fraysse, and
  Abergel]{HadjYoucef2018}
M.~e.~A. Hadj-Youcef, F.~Orieux, A.~Fraysse, and A.~Abergel, ``{Spatio-Spectral
  Multichannel Reconstruction from few Low-Resolution Multispectral Data},'' in
  \emph{Proc. European Signal Process. Conf. (EUSIPCO)}, Rome, Italy, Sept.
  2018.

\bibitem[{Space Telescope Science Institute
  (STScI)}(2017{\natexlab{b}})]{JDoxNIRSpecFilters}
\BIBentryALTinterwordspacing
{Space Telescope Science Institute (STScI)}, \emph{NIRSpec Dispersers and
  Filters, JWST User Documentation [Published 2017 March 29]}, Baltimore, MD,
  2017. [Online]. Available:
  \url{https://jwst-docs.stsci.edu/near-infrared-spectrograph/nirspec-instrumentation/nirspec-dispersers-and-filters}
\BIBentrySTDinterwordspacing

\bibitem[{Starck} and {Murtagh}(2006)]{Starck2006}
J.-L. {Starck} and F.~{Murtagh}, \emph{Astronomical Image and Data Analysis},
  ser. Astronomy and Astrophysics Library.\hskip 1em plus 0.5em minus
  0.4em\relax Springer-Verlag Berlin Heidelberg, 2006.

\bibitem[{Rauscher} et~al.(2007){Rauscher}, {Fox}, {Ferruit}, {Hill},
  {Waczynski}, {Wen}, {Xia-Serafino}, {Mott}, {Alexander}, and
  {Brambora}]{Rauscher2007}
B.~J. {Rauscher} \emph{et~al.}, ``Detectors for the {James Webb Space
  Telescope} near-infrared spectrograph. {I.} readout mode, noise model, and
  calibration considerations,'' \emph{Publ. Astron. Society of the Pacific},
  vol. 119, no. 857, pp. 768--786, Jul 2007.

\bibitem[{Space Telescope Science Institute
  (STScI)}(2017{\natexlab{c}})]{JDoxNIRSpecIFU}
\BIBentryALTinterwordspacing
{Space Telescope Science Institute (STScI)}, \emph{NIRSpec Integral Field Unit,
  JWST User Documentation [Published 2017 March 29]}, Baltimore, MD, 2017.
  [Online]. Available:
  \url{https://jwst-docs.stsci.edu/near-infrared-spectrograph/nirspec-instrumentation/nirspec-integral-field-unit}
\BIBentrySTDinterwordspacing

\bibitem[{Kelsall} et~al.(1998){Kelsall}, {Weiland}, {Franz}, {Reach},
  {Arendt}, {Dwek}, {Freudenreich}, {Hauser}, {Moseley}, and
  {Odegard}]{Kelsall1998}
T.~{Kelsall} \emph{et~al.}, ``The {COBE} diffuse infrared background experiment
  search for the cosmic infrared background. {II. Model} of the interplanetary
  dust cloud,'' \emph{Astrophys. J.}, vol. 508, no.~1, pp. 44--73, Nov 1998.

\bibitem[{Guilloteau} et~al.(2019){Guilloteau}, {Oberlin}, {Bern{\'e}}, and
  {Dobigeon}]{Guilloteau2019}
\BIBentryALTinterwordspacing
C.~{Guilloteau}, T.~{Oberlin}, O.~{Bern{\'e}}, and N.~{Dobigeon},
  ``Hyperspectral and multispectral image fusion under spectrally varying
  spatial blurs -- {A}pplication to high dimensional infrared astronomical
  imaging,'' \emph{ArXiv e-prints}, Dec. 2019. [Online]. Available:
  \url{http://arxiv.org/abs/1912.11868/}
\BIBentrySTDinterwordspacing

\end{thebibliography}

\end{document}